\documentclass[twocolumn, preprint2]{aastex61}
\usepackage{apjfonts}
\usepackage{url}
\usepackage{hyperref}
\usepackage{amsmath}

\usepackage{color}
\usepackage{comment}
\usepackage[normalem]{ulem}

\usepackage{longtable}
\usepackage[acronym, nogroupskip, sort=standard, nonumberlist, nopostdot]{glossaries}
\definecolor{dark-red}{rgb}{0.8,0,0}
\definecolor{dark-green}{rgb}{0,0.4,0}
\definecolor{dark-blue}{rgb}{0,0,0.8}
\definecolor{dark-margenta}{rgb}{0.8,0,0.8}
\definecolor{orange}{rgb}{1.0,0.6,0}
\definecolor{grey}{rgb}{0.6,0.6,0.6}
\newcommand{\fb}[1]{#1}
\newcommand{\stth}[1]{}

\DeclareMathOperator{\slog}{s-log}
\DeclareMathOperator{\sign}{sign}

\newcommand{\CH}[1]{${\rm CH}_{#1}$}
\newcommand{\SC}[1]{${\rm SC}_{#1}$}

\received{}
\revised{}
\accepted{}
\submitjournal{ApJ}
\shorttitle{2010 August 1--2 Sympathetic Eruptions: II.}
\shortauthors{Titov et al.}

\newcommand{\Rout}{R_{\rm out}}
\newacronym{hcs}{HCS}{heliospheric current sheet}
\newacronym{qsls}{QSLs}{quasi-separatrix layers}
\newacronym{bp}{BP}{bald patch}
\newacronym{bflp}{BFLP}{bracketing field-line pair}
\newacronym{chs}{CHs}{coronal holes}
\newacronym{pil}{PIL}{polarity inversion line}
\newacronym{dfrs}{DFRs}{disconnected flux regions}
\newacronym{scs}{SCs}{separatrix curtains}
\newacronym{dfss}{DFSS}{disconnected-flux separatrix surfaces}
\makeglossaries
\begin{document}

\title{2010 AUGUST 1--2 SYMPATHETIC ERUPTIONS: \\ 
II. MAGNETIC TOPOLOGY OF THE MHD BACKGROUND FIELD} 


\correspondingauthor{Viacheslav S. Titov}
\email{titovv@predsci.com}

\author{Viacheslav S. Titov}
\affiliation{Predictive Science Inc., 9990 Mesa Rim Road, Suite 170, San Diego, CA 92121} 

\author{Zoran Miki\'{c}}
\affiliation{Predictive Science Inc., 9990 Mesa Rim Road, Suite 170, San Diego, CA 92121} 

\author{Tibor T\"{o}r\"{o}k}
\affiliation{Predictive Science Inc., 9990 Mesa Rim Road, Suite 170, San Diego, CA 92121} 

\author{ Jon A. Linker}
\affiliation{Predictive Science Inc., 9990 Mesa Rim Road, Suite 170, San Diego, CA 92121} 

\author{Olga Panasenco}
\affiliation{Advanced Heliophysics, 1127 E Del Mar Blvd, Suite 425, Pasadena, CA 91106}

%\author{V. S. Titov\altaffilmark{1}, Z. Mikic\altaffilmark{1}, T. T\"{o}r\"{o}k\altaffilmark{1}, J. A. Linker\altaffilmark{1}}
%\and
%\author{O. Panasenco\altaffilmark{2}}
%
%\altaffiltext{1}{Predictive Science Inc., 9990 Mesa Rim Road, Suite 170, San Diego, CA 92121} 
%
%\altaffiltext{2}{Advanced Heliophysics, 1127 E Del Mar Blvd, Suite 425, Pasadena, CA 91106} 
%
%\email{titovv@predsci.com}

%\author{ } %\altaffilmark{2}

%\affil{ }

%\altaffiltext{1}{}
%\altaffiltext{2}{}

\begin{abstract}
\fb{Using a potential field source surface (PFSS) model,} we \fb{recently} analyzed the global topology of the background coronal magnetic field for a sequence of coronal mass ejections (CMEs) \fb{that occurred} on 2010 August 1--2. Here we repeat this analysis for the background field reproduced by a magnetohydrodynamic (MHD) model that incorporates plasma thermodynamics. As for the PFSS model, we find that \fb{all} three \fb{CME} source regions contain a coronal hole that is separated from neighboring coronal holes by topologically very similar pseudo-streamer structures. However, \fb{the two models yield very different results for the size, shape, and flux of the coronal holes. We find that the helmet-streamer cusp line, which corresponds to a source-surface null line in the PFSS model,} is structurally unstable and does not \fb{form} in the MHD model. Our analysis indicates that generally, in MHD configurations, \fb{this line rather consists of a multiple-null separator passing along the edge of disconnected flux regions. Some of these regions are transient} and may be the origin of so-called streamer blobs. We show that the core topological structure of such blobs is a three-dimensional ``plasmoid'', consisting of two conjoined flux ropes of opposite handedness, \fb{which connect at a spiral null point of the magnetic field}. Our analysis reveals that such plasmoids appear also in pseudo-streamers on much smaller scales. These new insights into the coronal magnetic topology provide some intriguing implications for solar energetic particle events and for the properties of the slow solar wind.
\\
\\
%The definition of the open separator field lines, along which interchange reconnect occurs, is also modified.
%Due to MHD flows, the basic nulls can bifurcate into several others to form three-dimensional ``plasmoids'' with spiral nulls at their centers.
%
\end{abstract}


\section{INTRODUCTION}
 \label{introduction}

Observations show that sequential coronal mass ejections (CMEs) can be widely separated in space but nearly synchronized in time, in which case a causal link between them is assumed to exist and so
 they are called sympathetic or linked.
To prove this hypothesis and identify underlying mechanisms of the link, in-depth investigations of particular events are required.
For example, \citet{Schrijver2011} combined SDO and STEREO observations to analyze a sequence of sympathetic CMEs that occurred on 1--2 August 2010, which sparked of other inspiring studies of these and other eruptions \citep{Schrijver2013, Jin2016}. 

Of particular interest to us is an idealized magnetohydrodynamic (MHD) simulation of the 2010 August 1-2 events \citep{Torok2011} that demonstrates that the topological structure of pseudo-streamers can precondition the linking and order of sequential eruptions.
Note that such a structure is primarily due to the background magnetic field rather than the field of erupting flux ropes.
Therefore, by analyzing just the background-field topology, one can gain essential insights into the linking of sympathetic eruptions that occur in pseudo-streamer configurations.
This is \stth{very} important from a technical point of view, as the \fb{stationary} background magnetic field is much easier to model \stth{by itself}\fb{on its own} than together with \stth{time-dependent} \fb{evolving} flux ropes.

Based on this premise, we have investigated in detail the field-line topology of a potential field source-surface 
\citep[PFSS,][]{Altschuler1969, Schatten1969} configuration that was reconstructed from a {\it Solar and Heliospheric Observatory}/MDI \citep{Scherrer1995} synoptic map available for the indicated time period \citep[][referred henceforth as Paper I]{Titov2012}.
This investigation confirmed the fundamentals of our idealized MHD simulation and provided new ideas on topological links of two additional eruptions that were not included in the simulation.

The advantage of the PFSS model is that it provides the simplest possible description of the global coronal magnetic field.
PFSS solutions often closely match MHD results for configurations based on line-of-sight magnetograms \citep{Riley2006}.
However, even in these cases, the PFSS model appears to be\stth{ inadequately} simplistic in the region close to the lower edge of the \gls{hcs}.
The latter is represented in the model by a source-surface null line, which is in general topologically unstable and has to disappear in a more realistic model of the global magnetic field.
Moreover, as\stth{ we already} \fb{was} pointed out in Paper I, some topological features, such as open separators, are sensitive to the presence of the null line at the source surface.
In our view, these features are critical for operating the so-called interchange reconnection between open and closed fields \citep{Crooker2002}, and hence for setting the\stth{ above-mentioned} causal link between sequential eruptions.
Thus, we arrive at the question of to what extent \stth{our} results obtained \stth{for} \fb{from a structural analysis of a}\stth{the} PFSS configuration\stth{ are actually} model-independent.

To \stth{answer} \fb{address} this question, we perform \fb{here} a similar \stth{structural} analysis \stth{as in Paper I,} \fb{by} using \stth{here} a full MHD model that \stth{includes the solar wind and} is derived from the same MDI data. 
Comparing \stth{the results obtained for} our MHD and PFSS models, we identify the salient similarities and differences in \stth{the} \fb{their} magnetic \fb{strucutures}\stth{topologies they provide}.
\stth{This allows us not only to answer the raised question but also to reassess the capability of the PFSS model to reproduce the global  structure of the coronal magnetic field.}

We start by describing the MHD model in Section \ref{s:model} and techniques for analyzing the field-line topology in Section \ref{s:tech}.
In Section \ref{s:magtop}, we present our results obtained \stth{by using} first \fb{from} only coronal holes' maps (Section \ref{s:hQl}) and then \fb{from} a combination of $Q$-maps and coronal holes' connectivity maps (Section \ref{s:cmaps});
the \fb{magnetic} topology of \fb{regions} \stth{with magnetic field}  \stth{lines} fully disconnected from the Sun and \stth{of the} pseudo-streamers are described in Section \ref{s:DFRs} and Section \ref{s:ssPS}, respectively.
In Section \ref{s:impl}, we \stth{address} \fb{consider} some ``spin-off'' implications of our analysis for coronal processes, and 
Section \ref{s:s} provides a summary and discussion of our work.

%%%%%%%%%%%%%%%%%%%%%%%%%%%%%%%%%%%%%%%%%%%%%%%%%
%%
%%                           Acronyms
%%
%%%%%%%%%%%%%%%%%%%%%%%%%%%%%%%%%%%%%%%%%%%%%%%%%

\fb{
We use the following acronyms for structural elements of magnetic configurations:
%\vspace{-10mm}
\vspace{-\baselineskip}
\printglossary[type=\acronymtype, title=
] %type=\acronymtype, title=\textbf{Acronyms for Structural Elements}
}

\section{INVESTIGATION METHODS} 
	\label{s:methods}

\stth{In }This new study of the coronal magnetic \stth{configuration} \fb{field at the time of}\stth{for} the 2010 August 1--2 sympathetic CMEs \fb{uses}\stth{basically} two methods\stth{ are used}.
First, we apply our numerical MHD model of the solar corona (Section \ref{s:model}) to reconstruct this \stth{configuration} \fb{field} from the observed magnetic data.
\stth{Then, }\fb{For} analyzing \stth{the resulting magnetic} \fb{its} structure, we \fb{then} employ\stth{ our} updated techniques (Section \ref{s:tech})\stth{that}\fb{, which} are based on a point-wise mapping of the boundar\fb{ies} \stth{surfaces} on themselves along magnetic field lines.
\stth{By this means we} \fb{These techniques enable us to} detect structural features of the magnetic field in the volume or at the boundaries and \fb{to} determine \fb{comprehensively} the related field-line topology.
\stth{The combination of these two methods enables us to investigate comprehensively the global magnetic structure, as given by our MHD model, and compare it with the results we have previously obtained for the PFSS model.}

\subsection{MHD Model } %Description
	\label{s:model}
	
In order to highlight the differences between the PFSS model and the MHD model, we used an identical photospheric magnetic field distribution \citep{Titov2012} \fb{as boundary conditions for} both models.
 The radial component of the photospheric magnetic field is based on a SOHO/MDI synoptic map of Carrington rotation 2099 (observed from 13 July to 9 August, 2010), suitably filtered, including smoothing and geometrical fitting of the polar fields.
\stth{For the calculation of the magnetic field we used} Our MHD model \citep{Lionello2009} \stth{with}\fb{employs} improved energy transport, including an empirical parameterized coronal heating specification, anisotropic thermal conduction along the magnetic field lines, radiative losses, and solar wind acceleration using a simplified WKB model of Alfv\'en wave propagation \citep{Jacques1977}.
\stth{We included} The upper chromosphere and transition region \fb{with their steep temperature and density gradients are included} in the model, \stth{by} starting at a temperature of $20{,}000\,{\rm K}$ at the base of our radial domain.
%, and \stth{includes}\fb{incorporates}  that form in the transition region.
This model \stth{describes}\fb{determines} the equilibration between open and closed magnetic field\fb{s} \stth{lines} in the corona, including the acceleration of the solar wind along open field lines to supersonic speeds.
%\stth{The model, as well as its application, is described in more detail by \citet{Lionello2009}.}
\stth{The description of} \fb{It is successfully} \stth{application}\fb{applied} to a coronal prediction for the August~1, 2008 total solar eclipse\stth{ is described} by \citet{Rusin2010}.

We used gas-characteristic boundary conditions parallel to the magnetic field lines at the lower radial boundary (\stth{which is in }the upper chromosphere), and outward characteristics at the upper radial boundary, where the flow is supersonic and super-Alfv\'enic.  The calculation was performed on a nonuniform spherical ($r$,$\theta$,$\phi$) mesh, using $252\times 368\times 661$ mesh points in a domain 
with the upper boundary at $r=20R_\sun\equiv\Rout$.
The radial resolution varied from $\Delta r=216\,{\rm km}$ at $r=R_\sun$, increasing to $\Delta r=300\,{\rm km}$ in the corona at $r\simeq 1.02R_\sun$, $\Delta r=800\,{\rm km}$ at $r\simeq 1.05R_\sun$, and $\Delta r=0.6R_\sun$ at $r\simeq 10R_\sun$.
In the ($\theta$,$\phi$) \stth{surface}\fb{plane}, the mesh points were concentrated in the region of interest, encompassing the principal active region and the two filament channels in the northern hemisphere (marked as 2', 2, and 3 in Figure \ref{f:Br+CHs}, respectively).
The smallest mesh cells, in the active region, were $0.18^\circ\,\times\,0.18^\circ$, gradually increasing to $0.37^\circ\,\times\,0.37^\circ$ in the region of the two quiescent filaments.  Far from the region of interest, the mesh cells increased to $\Delta \theta=2.6^\circ$ (at the poles), and $\Delta \phi=1.1^\circ$.  The calculation was performed on $3{,}360$ processors of NASA's Pleidas supercomputer.

Our thermodynamic MHD model uses an empirical coronal heating source $H$.  Over the years we have settled on two principal empirical heating sources (which we call Version 1 and Version 2).  Each heat source has a handful of parameters to link the heating to the magnetic field, the magnetic \stth{field} topology (via computed closed and open field regions), and to specified heating scale lengths.  We have invested considerable effort
in finding coronal heating specifications that produce realistic solar wind solutions, while simultaneously matching emission in EUV and X-rays \citep{Lionello2009}.
The model described in the present paper uses Version 2 of the heating model.
\stth{Incidentally, this same heating model was used in our prediction of the state of the corona prior to the July~11, 2010 total solar eclipse, which preceded the sympathetic eruptions by three weeks.  This prediction is described at \textcolor{blue}{\texttt{http://www.predsci.com}}.}
Briefly, the heating is \fb{defined by} the sum of four terms.  The first is an active-region and quiet-sun term in which the heating falls off exponentially from the solar surface with a scale height of $\lambda=0.06R_\sun$, with an amplitude that is proportional to $B_{\rm photo}$, the local magnitude of the photospheric magnetic field.  This heating is used only in closed-field regions, as estimated from a PFSS model.
The second term is a ``neutral line'' heating whose magnitude falls away from the photospheric \gls{pil} \fb{at which} $B_{r\:{\rm photo}}=0$.  This heating has a scale length $\lambda=0.03R_\sun$, and is also proportional to $B_{\rm photo}$.  The third term is a ``fast wind'' heating with a scale length $\lambda=1.0R_\sun$ that is applied everywhere.  The fourth term is a ``short-scale'' heating with a scale length $\lambda=0.03R_\sun$ that is also applied everywhere.
This {\it ad hoc} feature of the heating model arises from the complex nature of coronal heating, a subject that is currently not well understood.  In practice, our heating specification produces solutions with reasonable coronal properties.  A comparison between an MHD simulation employing this heating model and ground-based and spacecraft observations of the July~11, 2010 total solar eclipse showed good agreement, including the locations of coronal holes and coronal streamers, and brightness in X-rays and EUV.

The model was relaxed for 55 hours, when the magnetic field reached a state of quasi-equilibrium.  The equilibration between open and closed magnetic field structures was achieved by this time.  Some of the low-lying loops in the active region, where the heating was very strong, experienced thermal nonequilibrium throughout the simulation, with cyclic condensations and evaporations of plasma \citep[e.g.,][]{Mikic2013c}.  In this sense, for the parameters selected, there is no strict steady state in the MHD model, a common occurrence when the plasma is heated to observed coronal temperatures \citep{Mok2016}.  Nevertheless, the large-scale coronal structures are in a state that is close to equilibrium.

\subsection{Techniques for Analyzing the Structural Skeleton of Magnetic Configurations}
	\label{s:tech}

The {\it structural skeleton} of a magnetic configuration is formed by separatrix surfaces and \gls{qsls} \citep[][]{Priest1995, Demoulin1996, Titov1999a}.
Its determination for a realistic magnetic field model is a very important but generally challenging task.
Significant progress in solving this problem has been made by developing automated procedures for calculating the skeleton footprints at the boundaries of computation domains and at their cross-sections  \citep{Titov2008a, Pariat2012, Liu2016, Tassev2016}.
These footprints are identified as two-dimensional regions characterized by high values of the so-called {\it squashing degree or factor} $Q$ of elemental magnetic flux tubes \citep{Titov1999a,Titov2002,Titov2007a}.
$Q$ is defined at both footpoints of a given field line and, by construction, has the same value there.
It measures the divergence of the neighboring, infinitesimally close, field lines from the given field line.
Typical $Q$ distributions at the boundaries have sharp ridges with very narrow bases, which are called hereafter high-$Q$ lines or curves.
So the structural skeleton can, in principle, be retrieved by tracing appropriate field lines from the high-$Q$ curves.
Thus, our general approach to analyze the magnetic structure is as follows: We first calculate the required $Q$-maps and then trace a set of field lines whose footpoints best localize the $Q$-ridges on these maps.
The first part of this approach is already well developed, while the second one is new and rather nontrivial.
We discuss it in detail below in Section \ref{ss:BFLP} and Section \ref{ss:CHCM}.

\subsubsection{The Structural Skeleton in Terms of $Q$-maps}

Since both boundaries in our magnetic-field model are spheres,
we compute the $Q$-maps at the boundaries by using the expression of $Q$ in terms of the Jacobi matrix elements for the field-line mapping determined in spherical latitude-longitude coordinates $(\theta,\phi)$ \citep{Titov2007a, Titov2008a}.
For interpreting these maps, we find it useful to show on them additionally the sign of the normal field $B_{r}$ at the boundary.
It is convenient to do this by using the function called {\it  signed log Q}, or simply $\slog Q$, defined as \citep{Titov2011}
\begin{eqnarray}
  \slog Q \equiv \sign(B_r) \log\left[Q/2 + \left(Q^2/4-1\right)^{1/2}\right] \,.
	\label{slogQ}
\end{eqnarray}
Using a two-color scale bar that is symmetric about zero for plotting $\slog Q$ distribution, we automatically show sign of $B_{r}$ on the $Q$-map.
Letting these colors rapidly fade from the ends to the middle of the color bar,
one also sharpens high-$Q$ lines and thereby improves the readability of complex $\slog Q$-maps.
We employ here such a sharpening aqua-crimson palette (see Figure \ref{f:slogQ_maps}) instead of the blue-red palette that we used earlier in Paper I.
As before, it is also helpful to combine this color coding with a transparency mask in which the colors are made fully transparent at $Q \lesssim 10^2$ and gradually becoming fully opaque at, say, $Q \approx 10^3$.
Thus prepared, the $\slog Q$-maps can be superimposed on corresponding $B_{r}$-maps without much obscuring each other, even if the $B_{r}$-maps themselves are colored in blue-red (see, e.g., Figures \ref{f:DFSSvsSC} and \ref{f:SC1} below).

As mentioned above, both footpoints of a field line have the same value of $Q$, which therefore can be assigned to the field line itself, as a whole.
The $Q$ factor thus becomes extended from the boundary into the volume.
This formally implies that $Q$ has to satisfy the equation
\begin{eqnarray*}
   {\bf B \cdot \nabla} Q = 0 \,,
\end{eqnarray*}
where ${\bf B}$ is assumed to be known in the volume.
With this extension, we can define the structural skeleton as a high-$Q$ subvolume, which is essentially a set of QSLs.
The computation of $Q$-maps at the cross-sectional planes in addition to those at the boundaries facilitates understanding the structural skeleton \citep{Aulanier2005, Titov2008a, Pariat2012, Savcheva2012b, Savcheva2012}.
\stth{However, even with the help of this powerful technique, it remains difficult to interpret the $Q$-maps of realistic configurations  due to their sheer complexity.}

\subsubsection{The BFLP Method for Calculating (Quasi-)Separaratrix Surfaces}
\label{ss:BFLP}

\fb{However, even with the help of} \stth{this}\fb{the described powerful technique, it remains difficult to interpret the $Q$-maps of realistic configurations, due to their sheer complexity.}
To mitigate this problem, we have developed a new method that allows one to calculate efficiently the field-line structure of the (quasi-)separatrix surfaces by starting from their high-$Q$ footprints.
The underlying algorithm iteratively determines sets of field-line pairs that bracket the  structural features, such as null, minimum \citep{Titov2009}, and \gls{bp} \citep[][]{Titov1993, Seehafer1986a} points, which give birth to the (quasi-)separatrix surfaces.
Convergence of the algorithm towards the high-$Q$ footprint from either side automatically approximates these surfaces and the field-line structures at the related features. 

A (quasi-)separatrix field line is approximated at each iteration by first tracing many field lines whose footpoints are located between those of the present \gls{bflp}, then selecting from them a new BFLP (the pair of field lines with the highest divergence) for the next iteration.
It is clear that this method is less efficient as the one that traces separatrix field lines from points that are preliminary obtained by  analyzing null-point neighborhoods \citep{Haynes2010}. 
However, while loosing in efficiency, our BFLP method has the advantage to remain viable even when the numerical field data are not suitable for the null-neighborhood analysis.
Such a situation arises when the numerical resolution is not high enough in a region where a null point is predicted to be present by, say, a PFSS model.
Indeed, for sufficiently strong MHD perturbations of such a region, the neighborhood of the expected null point can collapse into a current layer of a thickness that is comparable with or even smaller than the local size of the grid.
This might cause the null-neighborhood analysis to fail, as it relies on the matrix of magnetic-field gradients, which tends to degenerate in such a situation.
In contrast, our BFLP method then still produces qualitatively sensible results.

\subsubsection{The Coronal-Holes' Connectivity Maps}
\label{ss:CHCM}

The high-$Q$ footprints of the (quasi-)separatrix surfaces at the upper boundary are related to helmet streamers and pseudo-streamers.
For an overall assessment of the global magnetic structure, it is useful to combine these maps with what we call \fb{connectivity maps of \gls{chs}}.
The latter depict the photospheric CHs along with their boundary field lines shown in projection to the photosphere \citep{Liu2006, Liu2007b}.
We have refined this type of visualization in such a way that the CH-boundary field lines rooted in opposite photospheric polarities  meet pairwise at the upper boundary (see Figure \ref{f:CH_conn_maps} below).
It is difficult to make this refinement straightforwardly, i.e., by adjusting their starting footpoints at the photosphere.
Fortunately, the desirable result is achieved also by simply using the upper-boundary footpoints as starting ones.
For technical reasons (see Section \ref{s:magtop}), it makes sense to choose an upper boundary at $r=\Rout^{*} < \Rout$ for analyzing the coronal magnetic structure determined by the MHD model in the domain between $R_{\sun}$ and $\Rout$.
Then the field lines of interest are those closed field lines which attain the upper boundary at the PIL defined by the equation $B_{r}(\Rout^{*},\theta, \phi)=0$.
Depending on the used model of the global configuration, 
either all or a part of the solutions of this equation determine the points where the CH-boundary field lines meet.
Starting from these points, we can trace our field lines downwards to the borders of the photospheric CHs.
Then projecting these field lines to the photosphere yields the  refined CHs' connectivity map. 

In configurations derived from MHD models, the magnetic field is generally nonzero along the upper-boundary PIL, which implies that each of its points is passed through by a unique field line.
However, only those of these field lines that approach the PIL from below are connected to the photospheric borders of CHs.
In other words, only the upper-boundary BPs (see the BP criterion for spherical boundary in Appendix \ref{s:bpc}) should be used for initializing CH-boundary field lines.
Tracing these field lines from the PIL forward (backward), we determine how the corresponding side of the HCS boundary is connected to a negative (positive) polarity at the photosphere.
The remaining segments of the upper-boundary PIL belong to \gls{dfrs} (see Section \ref{s:DFRs}).

For PFSS models, the described approach should be modified, because the magnetic field vanishes at the  source-surface PIL.
The latter turns in this case into a null line at which the uniqueness of passing field lines is no longer valid.
However, analyzing the field-line structure near this null line, we derive certain small displacements from it to non-null points of the helmet-streamer separatrix dome (Appendix \ref{s:HSSDob}).
Then, \stth{using}\fb{by tracing field lines from} these points\stth{ as starting ones}, the dome is uniquely recovered\stth{ by field-line tracing}.
%Then, using these points as starting ones, the dome is uniquely recovered by field-line tracing.
 
In the CHs' connectivity maps thus created for both of the models, it is instructive to take the same individual color for each CH and its boundary field lines.
The use of, say, reddish and bluish tints helps also to distinguish between CHs of positive and negative polarities, respectively.
This color scheme significantly enhances the informational content of the maps, as it
enables one to show clearly how individual CHs of opposite polarities get into contact with one another, and how their boundary magnetic fluxes split at the HCS and close onto one another (Figure \ref{f:CH_conn_maps}).

In addition, \fb{by} superimposing upper-boundary $\slog Q$-maps onto related CHs' connectivity maps, we gain two \stth{additional}\fb{extra} capabilities.
First, this superimposition allows us to identify \fb{on the HCS the exact locations} where the CH-boundary fluxes split.
Second, it elucidates the relationship of the upper-boundary footprints of the pseudo-streamer \stth{separatrix curtains}\fb{SCs} to the associated (basic) null points low in the corona.
Some of the boundary field lines in two neighboring CHs of like polarities encounter at these points and diverge before reaching the upper boundary.
As shown in Section \ref{s:cmaps}, counter-streaming patterns of these field lines clearly reveal the locations of the basic nulls on the maps.
% highly divergent character of the counter-streaming field lines towards these points.

%%%%%%%%%%%%%%%%%%%%%%%%%%%%%%%%%%%%%%%%%%%%%%%%%
%%
%%                      MAGNETIC FIELD STRUCTURE
%%
%%%%%%%%%%%%%%%%%%%%%%%%%%%%%%%%%%%%%%%%%%%%%%%%%

\section{MHD VS. PFSS: COMPARATIVE STUDY OF THE MAGNETIC STRUCTURE}
	\label{s:magtop}

PFSS configurations have only two types of magnetic field lines: (1) closed field lines with both footpoints at the photosphere and (2) open field lines with one footpoint at the photosphere and the other at the upper boundary\fb{,} typically chosen at $r=\Rout=2.5\, R_{\sun}$\stth{ called} \fb{(the source surface)}.
To account for the field-line stretching by the solar wind, the upper-boundary condition is \stth{chosen}\fb{prescribed such that}\stth{ so as to force} the\stth{ local} \fb{computed source-surface} field\stth{ to be strictly} \fb{would be fully} radial.
As mentioned above, this boundary condition automatically nullifies the field at the source-surface PIL.
The PFSS model assumes that the HCS is rooted at this line, which tacitly implies that the field vanishes all over the HCS midsurface as well.
\stth{However, the thus inferred magnetic structure of the HCS, as shown below, is oversimplified in comparison to what our more realistic MHD model yields.}
\fb{However, a comparison with more realistic MHD model shows that the thus inferred structure of the HCS is oversimplified}.

In particular, the MHD-modeled HCS has neither a null line nor a null surface\stth{ at all}, implying that those are \stth{both PFSS-model} artifacts \fb{of the PFSS model}.
Due to a free-outflow boundary condition imposed in MHD at the uppermost radius $\Rout$,
the magnetic field in the HCS has enough freedom to be almost, but not fully, radial.
As inferred below, the HCS must have, in general, a separator field line threading several null points, instead of the source-surface null line.
This feature turns out to be inherently related to the presence of DFRs, which consist of field lines
whose both footpoints are located at the upper boundary.

In the following, we will restrict our topological analysis of the MHD model mainly to the domain $R_{\sun}\le r\lesssim  5\, R_{\sun} (\equiv \Rout^{*})$.
This can be done almost without a loss of information on the field-line opening,
since the bulk of the field lines becomes straightened out by the solar wind already at $ r \approx 4\,R_{\sun}$.
The restriction to $5\, R_{\sun}$ also significantly speeds up field-line tracings, which are required in vast numbers for our analysis.

\subsection{Coronal-Holes' Maps}
	\label{s:hQl}

\begin{figure*}[ht]
%  \begin{comment}
\epsscale{1.1}
%\plotone{fr_afuh}\\
\plotone{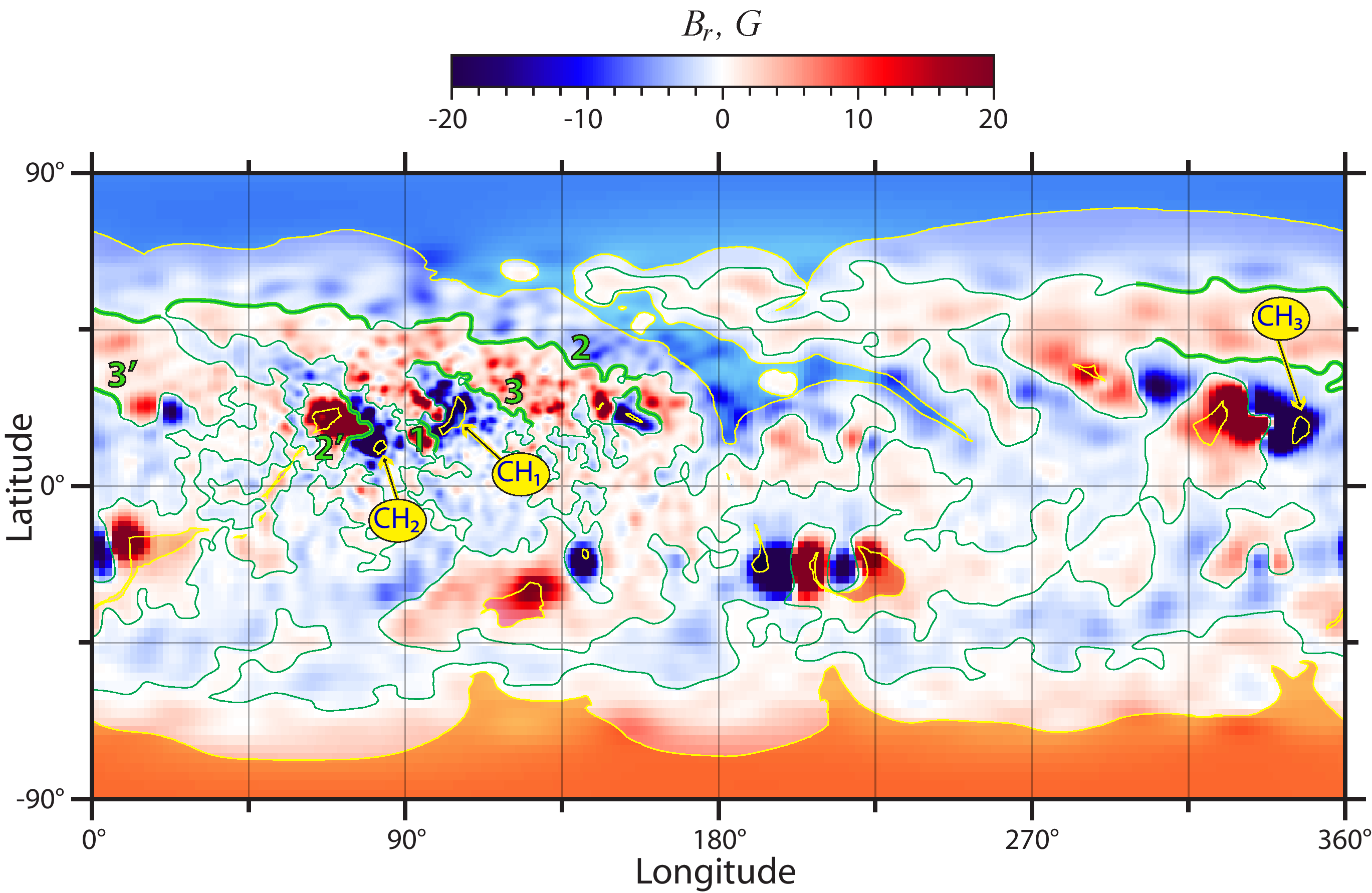}\\
\vspace{5pt}
\plotone{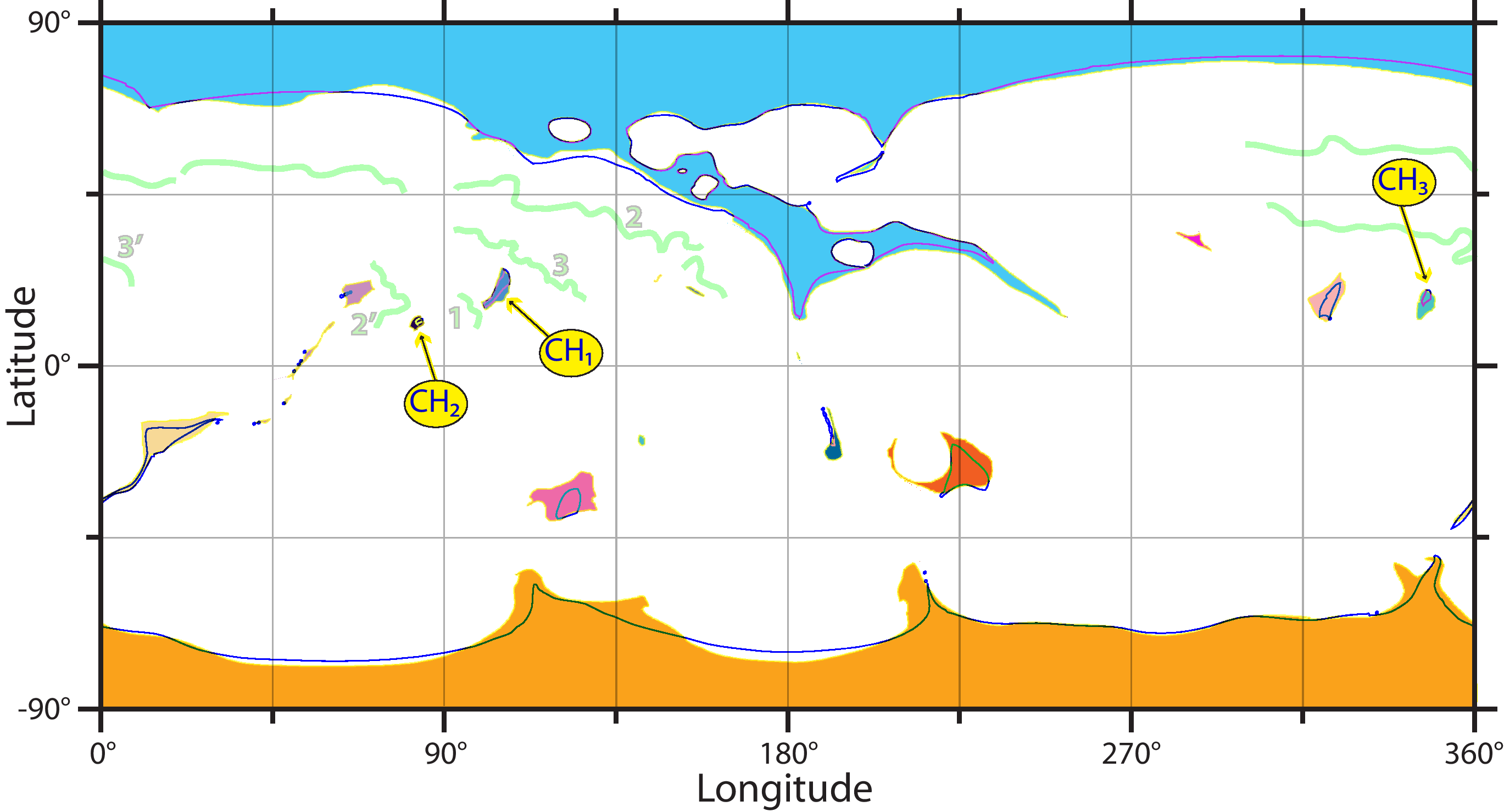}\\
%\plotone{CH_FL_conn_map_thermo_MHD}\\

%   \end{comment}
%\ \\
\caption
{
Top panel: superimposed photospheric maps of $B_{r}$ and coronal holes (semi-transparent colors, boundaries outlined in yellow) in our MHD model of the 2010 August 1--2 solar corona.
Green lines represent the PIL at $r=1.05\, R_{\sun}$,
with their thick segments designating the location of filaments, some of which are numbered in the order they erupted
%\footnote{This figure has the same PIL as in \protect\citet{Titov2012}, where it was misprinted as photospheric, i.e., as determined at $r=R_{\sun}$.}.
%\protect
\tablenotemark{a}. 
Bottom panel: comparison between MHD (shaded) and PFSS (contoured) coronal holes; colors with bluish and reddish tints correspond to negative and positive polarities, respectively.
Yellow balloons indicate in both panels the coronal holes (CH$_{1\ldots 3}$) involved in the eruptions.\\
\\
{\footnotesize $^{\rm a}$\ This figure has the same PIL as in \citet{Titov2012}, where it was misprinted as photospheric, i.e., as determined at $r=R_{\sun}$}.
}
	\label{f:Br+CHs}
\end{figure*}

The photospheric \stth{coronal holes'}\fb{CHs'} map\stth{ has been} \fb{is} computed on a uniform grid with an angular cell size of $0.125\arcdeg$---the same we used for our PFSS configuration.
Overall, both models' maps give similar results, particularly for the source region of the observed CME sequence (top panel in Figure \ref{f:Br+CHs}).
As in our PFSS model, this region contains three coronal holes of negative polarity (${\rm CH}_{i},\  i=1,2,3$) that are clearly isolated from one another by positive parasitic polarities.
However, they are now significantly larger in size and different in shape, particularly ${\rm CH}_{2}$ and ${\rm CH}_{3}$.
Even more dramatic changes are seen for the \stth{coronal hole}\fb{CH} formed in the positive polarity of the active region close to filament $2^{\prime}$: being barely noticeable in the PFSS model, it turns\stth{ in the MHD model} into a \stth{coronal hole}\fb{CH}\stth{ whose size is more than} twice as large as \stth{that of} ${\rm CH}_{2}$.

These differences are likely systematic in origin,
because all low-latitude coronal holes show predominantly similar differences between the two models (bottom panel in Figure \ref{f:Br+CHs}).
The tendency of the MHD model to produce a larger open-field area also manifests in the presence of
several smaller coronal holes that have no analogues in the PFSS model.
Only the polar coronal holes of the two models could be regarded as having almost identital properties.

%{\hR
%Total open flux???
%}

Thus\stth{, in comparison to the PFSS model,} the MHD model yields\stth{ at low latitudes} a more open and complex structure \fb{at the low-latitude photosphere, which corresponds in the high corona}
\stth{, this photospheric property translates into} \fb{to} a wider equatorial belt of (quasi-)separatrix surfaces \citep[the so-called S-web, ][see also Section \ref{s:cmaps}]{Antiochos2011, Linker2011, Titov2011} 

Of particular importance is the fact that\stth{ the transition to} the MHD model makes our source-region \stth{coronal holes}\fb{CHs larger}\stth{ more prominent by increasing them in size}.
As will become clearer below, this \stth{provides an additional substantiation of} \fb{additionally substantiates} 
our earlier explanation for the causal link between sympathetic CMEs through pseudo-streamer topological structures.

\subsection{$\slog Q$-Maps and Coronal-Holes' Connectivity Maps}
	\label{s:cmaps}

\begin{figure*}[ht]
%  \begin{comment}
\epsscale{0.4}
\plotone{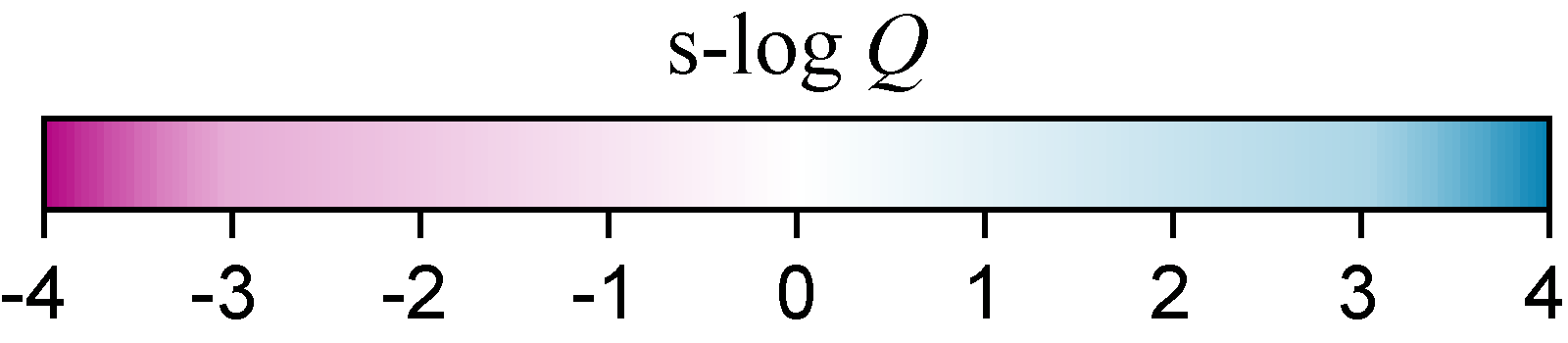}\\
\vspace{5pt}
\epsscale{1.1}
\plotone{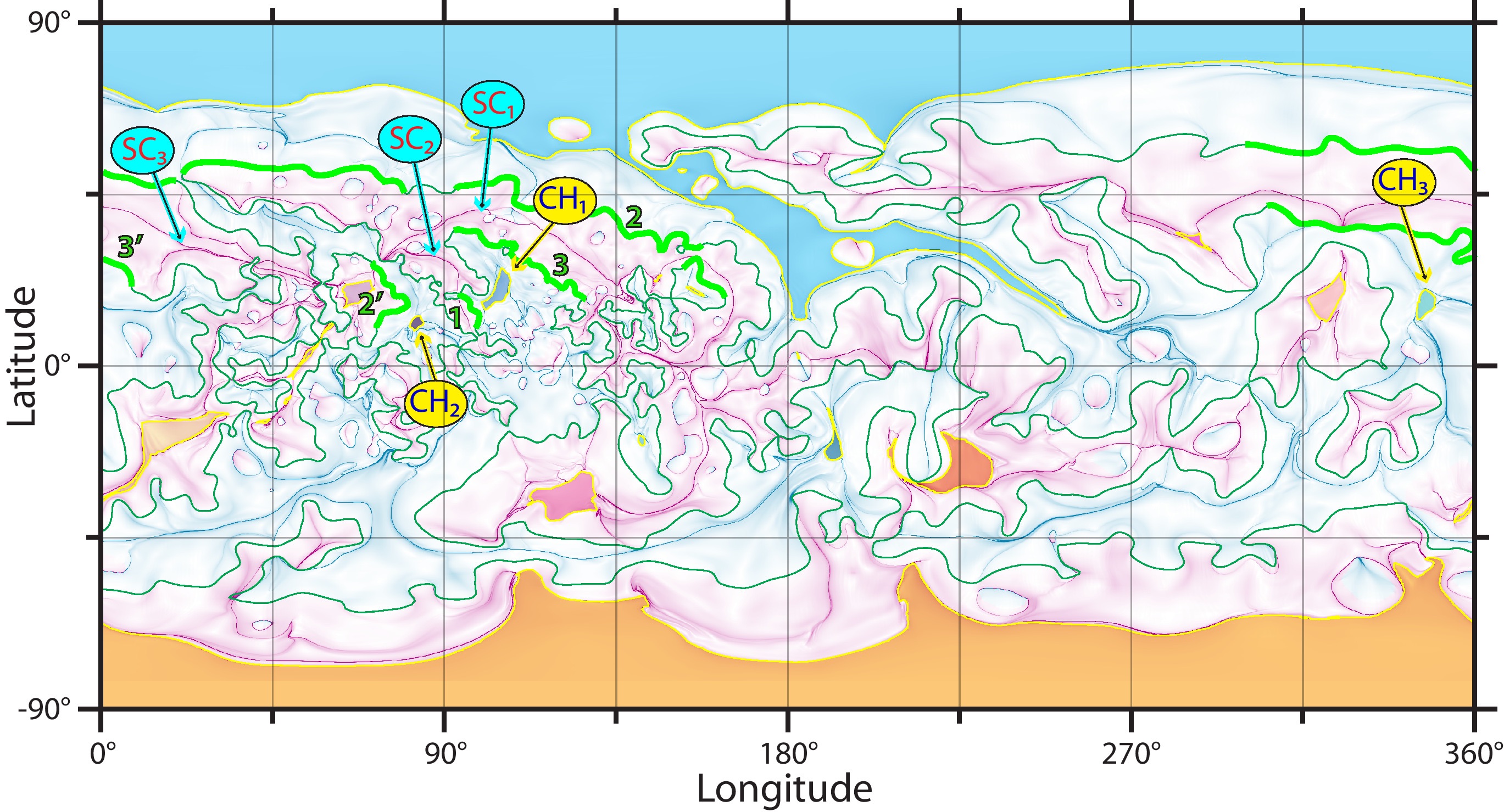}\\
\vspace{5pt}
\epsscale{1.1}
\plotone{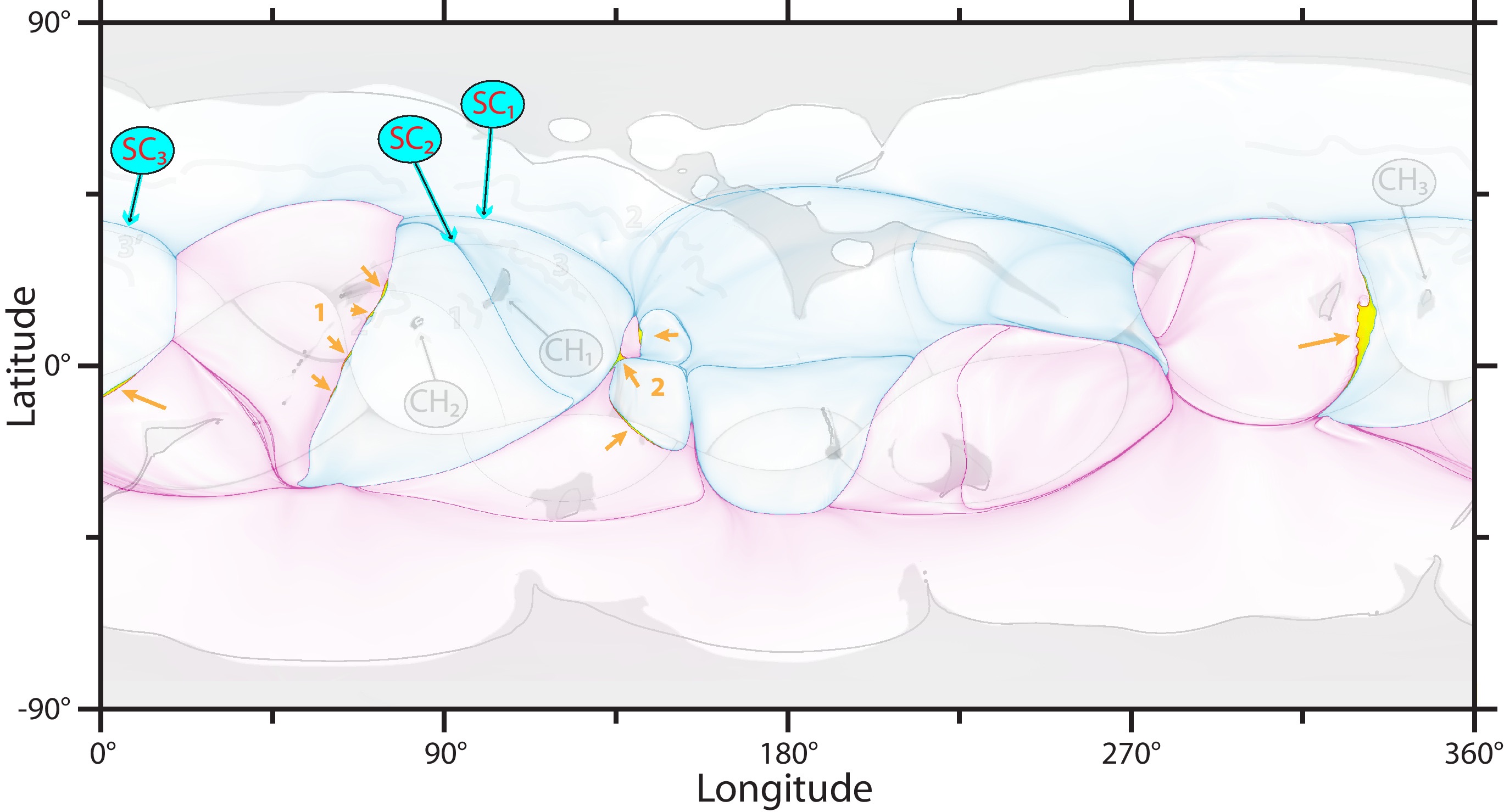}\\
%   \end{comment}
%\ \\
\caption
{
$\slog Q$ map of the MHD model at the photosphere (top panel) and upper boundary $r\approx 5\, R_{\sun}$ (bottom panel).  As in Figure \ref{f:Br+CHs}, the photospheric coronal holes in negative and positive polarities are shaded in colors with bluish and reddish tints, respectively.
The orange arrows point to the yellow-shaded areas of disconnected-flux regions in the upper-boundary map, which is superimposed on dimmed maps, including a similar $\slog Q$ map of the PFSS model and the photospheric coronal-holes' map shown in the bottom panel of Figure \ref{f:Br+CHs}.   The upper-boundary footprints of the pseudo-streamer separatrix curtains (SC$_{1\ldots 3}$) involved in the sympathetic eruptions are indicated with cyan balloons.
}
	\label{f:slogQ_maps}
\end{figure*}

To gain \stth{a better}\fb{further} insight\fb{s} into \stth{our magnetic}\fb{the} structure \fb{of our configuration}, let us consider its $\slog Q$ maps at the photosphere and the upper boundary $r=\Rout^{*}$.
Both maps are presented in the top and bottom panels of Figure \ref{f:slogQ_maps}, respectively.
Comparing the top panel with Figure 3 in Paper I, we see that they are very similar, except near the low-latitude coronal holes discussed above.

In particular, three prominent high-$Q$ lines, labeled in Figure \ref{f:slogQ_maps} (top panel) as ${\rm SC}_{i},\ i=1,2,3,$ are identified in the region of interest.
They are photospheric footprints of the \gls{scs} we described in Paper I.
Each of these SCs bisects the associated parasitic polarity and serves as an interface between two adjacent arcades of closed magnetic field rooted in that polarity.
Thus, in the low corona, the basic structure of the pseudo-streamers in the two models is very similar (see more details in Section \ref{s:ssPS}).

The situation, however, dramatically changes higher up in the corona, where the isolated \fb{low-latitude} CHs \stth{at low latitudes} rapidly expand toward the upper boundary.
As their expansions are rather anisotropic and \stth{individual}\fb{nonuniform}, their cross-sections become strongly distorted with increasing radius.
\stth{Their resulting shapes}\fb{They are} outlined on the $Q$-map at $r=\Rout^{*}$ by high-$Q$ lines\fb{, forming} \stth{are organized in} a network of closely packed cells (i.e., an S-web; see bottom panel in Figure \ref{f:slogQ_maps}).
These cells are typically different from and larger than their PFSS analogues.
To facilitate comparison between the two models, the corresponding source-surface $Q$-map
is shown dimmed in the panel (for a full-colored version of that source-surface $\slog Q$-map, see Figure 2b in Paper I). 
One can see that, in contrast to an eye-like form of cells in the PFSS model, either oval or curvilinear polygonal cells are present in the MHD model.

As evident from the panel, there are also essential differences between the two models regarding the HCS.
Irrespective of the model used, the HCS structure is represented at the upper boundary by two adjacent high-$Q$ lines following the PIL; they are colored in aqua and crimson on our $\slog Q$-map.
In the PFSS model, these two lines are merged together all over the PIL, which is in that case a simple and smooth curve resembling a sinusoid.
In the MHD model, the PIL turns out to be a much more complex curve having several cusps and folds.
%It is also formed by a similar merging of 
The two high-$Q$ lines here are also merged together along the PIL, but not everywhere.
There are several segments of the PIL along which these lines are separated by narrow pod-like areas;  in the top panel, those are shaded in yellow.
% while along other segments they remain separated by enclosing narrow areas, which are shaded in the above panel in yellow.
These areas are cross-sections of the\stth{ above-mentioned} DFRs, which \fb{are} form\fb{ed} inside the HCS at $r\gtrsim 3 R_{\sun}$.
%Such regions, of course, are not present at all in the PFSS model, as it describes only the structure below the HCS in potential field approximation.
We will discuss these regions in more detail in Section \ref{s:DFRs}. 

\begin{figure*}[ht]
%  \begin{comment}
\epsscale{1.1}
%\plotone{fr_afuh}\\
\plotone{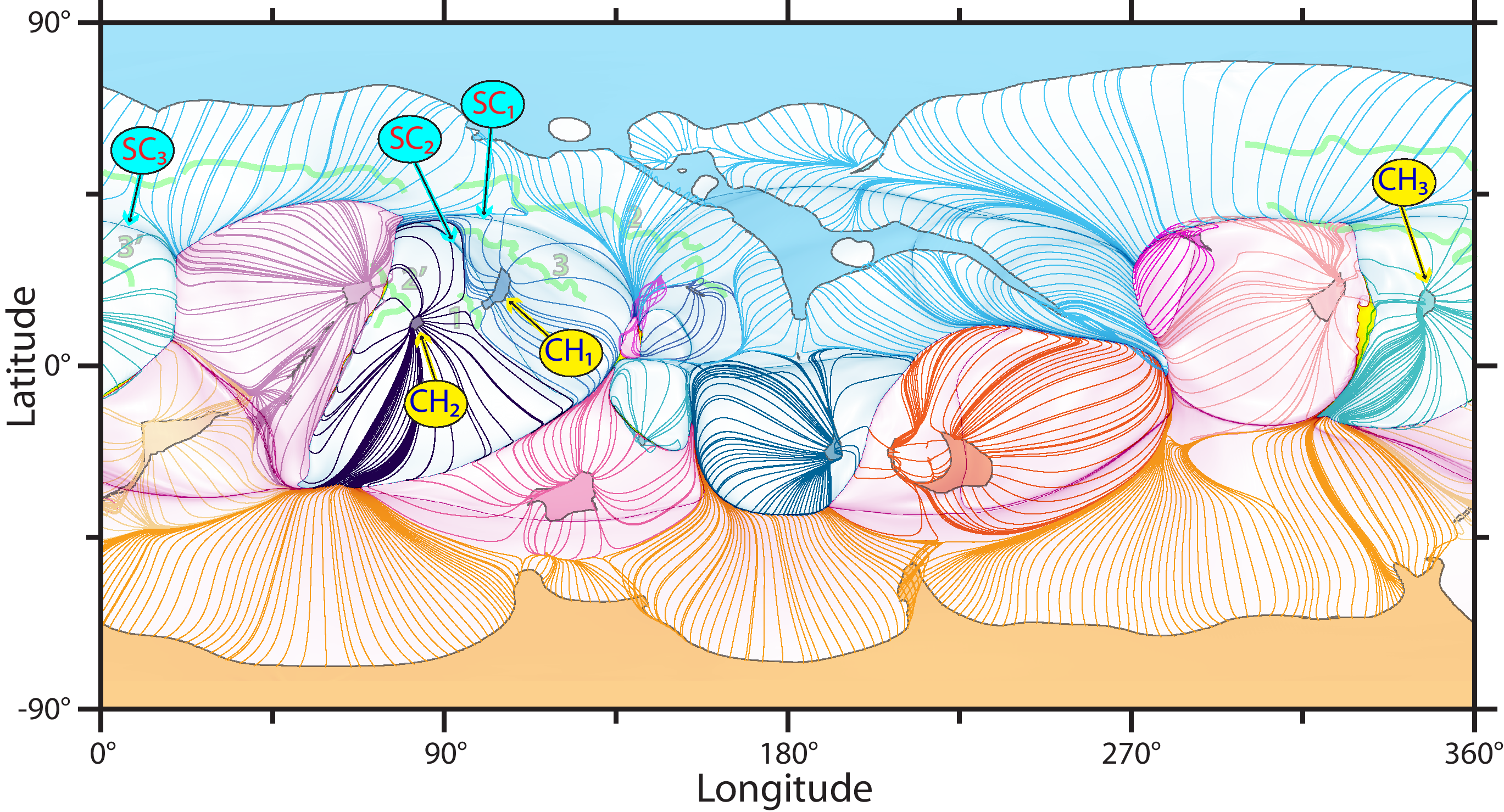}\\
\vspace{5pt}
\plotone{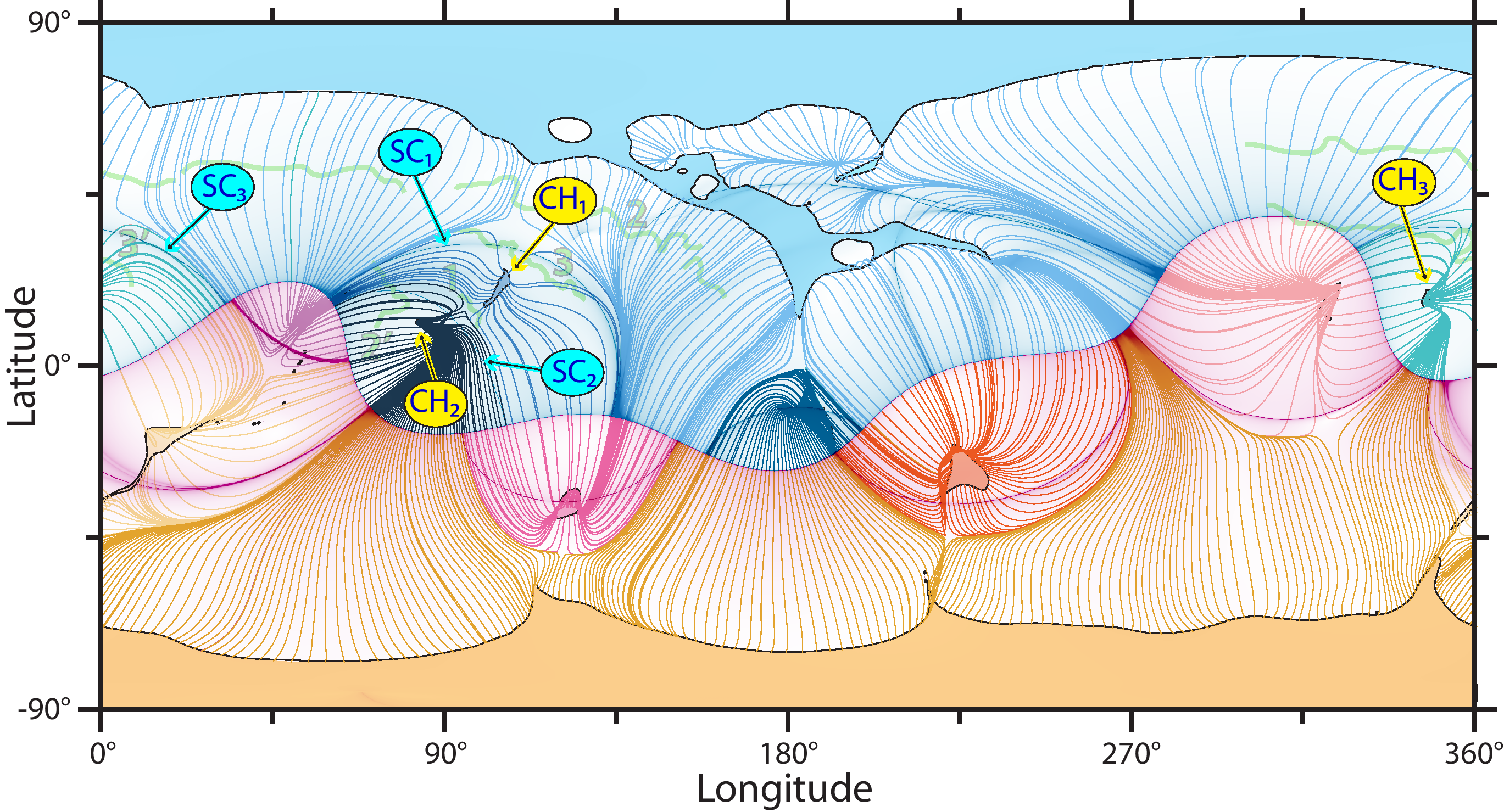}\\
%   \end{comment}
%\ \\
\caption
{
Coronal-holes' connectivity maps for MHD (top panel) and PFSS (bottom panel) models with superimposed semi-transparent $\slog Q$ maps at the upper boundaries $r\approx 5\, R_{\sun}$ and $r=2.5\, R_{\sun}$ for MHD and PFSS model, respectively.
Thick green lines show the location of filaments, some of which are numbered in the order they erupted.
The structural elements adjacent to them are indicated with yellow and cyan balloons, which
refer, respectively, to the coronal holes (\CH{1\ldots3}) and the upper-boundary footprints of the pseudo-streamer separatrix curtains (\SC{1\ldots3}).
}
	\label{f:CH_conn_maps}
\end{figure*}

The differences in magnetic structure between the two models become even more prominent in the CHs' connectivity maps.
As can be seen in  Figure \ref{f:CH_conn_maps}, this particularly concerns the low-latitude CHs:
Their boundary fluxes split at the HCS and close onto their counterparts in the opposite polarities quite differently in the models.
In particular, the CH-boundary fluxes from the polar CHs are fully ``intercepted'' at the HCS in the MHD model by low-latitude CHs of the opposite polarity.
However, in the PFSS model, this ``interception'' is only partial, so that a significant fraction of the polar CH-boundary fluxes remains closed onto one another.

According to these maps, the CH-boundary field lines behave differently also inside the HCS.
In the PFSS model (bottom panel in Figure \ref{f:CH_conn_maps}), they reach the source surface perpendicularly to the PIL -- null line,
%, which is a null line in this case, as we already know.
and conceivably continue above it radially and antiparallel across the HCS.
In contrast, in the MHD model most of such field lines reach the upper boundary by touching it at different angles with respect to the PIL. 
This is because only the radial magnetic field vanishes there, while the other components remain finite.
This is consistent with the top panel in Figure \ref{f:CH_conn_maps}, which implies that the field lines are not radial and antiparallel across the HCS but rather sheared along it.
Further evidence\stth{ on} \fb{for} this fact will be revealed in Section \ref{s:ssPS} from analyzing the field-line structure of the separatrix curtain ${\rm SC}_{1}$.

All these differences\stth{ in the global magnetic structure} \fb{between}\stth{ of} the\stth{ two} models are of \fb{a} general nature and \stth{likely to}\fb{therefore should} be\stth{ found} \fb{present} in other case studies as well.
Their full implications for coronal and heliospheric investigations have yet to be \fb{realized}.
At present, it seems that these differences\stth{ have significantly to} affect the models' predictions regarding the transport and escape of solar energetic particles from source regions located near CH boundaries (see Section \ref{s:impl}).

It is remarkable that, in spite of all these differences, the MHD model still firmly supports our previous results \citep{Torok2011, Titov2012} concerning the role of pseudo-streamers in the sympathetic eruptions on 2010 August 1--2.
Indeed, Figures \ref{f:slogQ_maps} and \ref{f:CH_conn_maps} suggest (and Section \ref{s:ssPS}  substantiates it) that, as in our PFSS model, there are three pseudo-streamers involved in these eruptions.
Their separatrix curtains ${\rm SC}_{i}$ higher up in the corona serve as boundaries between the coronal holes ${\rm CH}_{i}$ and either the northern polar CH (for $i=1,3$) or ${\rm CH}_{1}$ (for $i=2$), all of negative polarity.
Low in the corona, these three SCs continue down to positive parasitic polarities, to divide their closed-field structures into two lobes. Four of these six lobes contained the erupting filaments (all numbered ones except for $2^{\prime}$ in Figure \ref{f:slogQ_maps}) in a manner similar to what we found for the PFSS model.
This implies that the plausible scenario of sympathetic eruptions that we previously inferred in Paper I from our PFSS and idealized MHD simulation \citep{Torok2011} remains valid.

\subsection{Disconnected-Flux Regions and \\ HCS Magnetic Topology}
	\label{s:DFRs}

\begin{figure*}[ht]
%  \begin{comment}
\epsscale{1.15}
%\plotone{fr_afuh}\\
\plotone{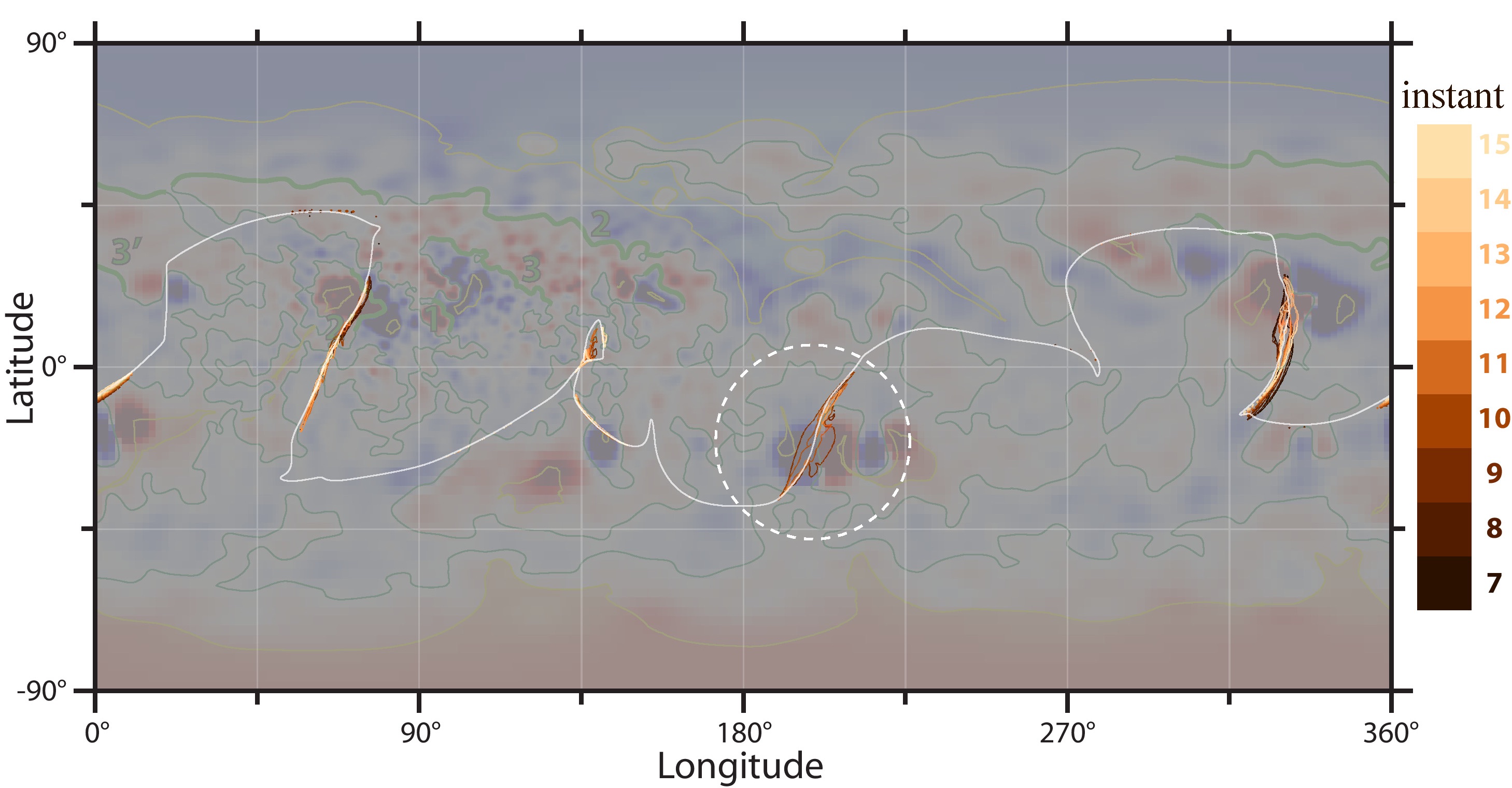}\\
\vspace{5pt}
\plotone{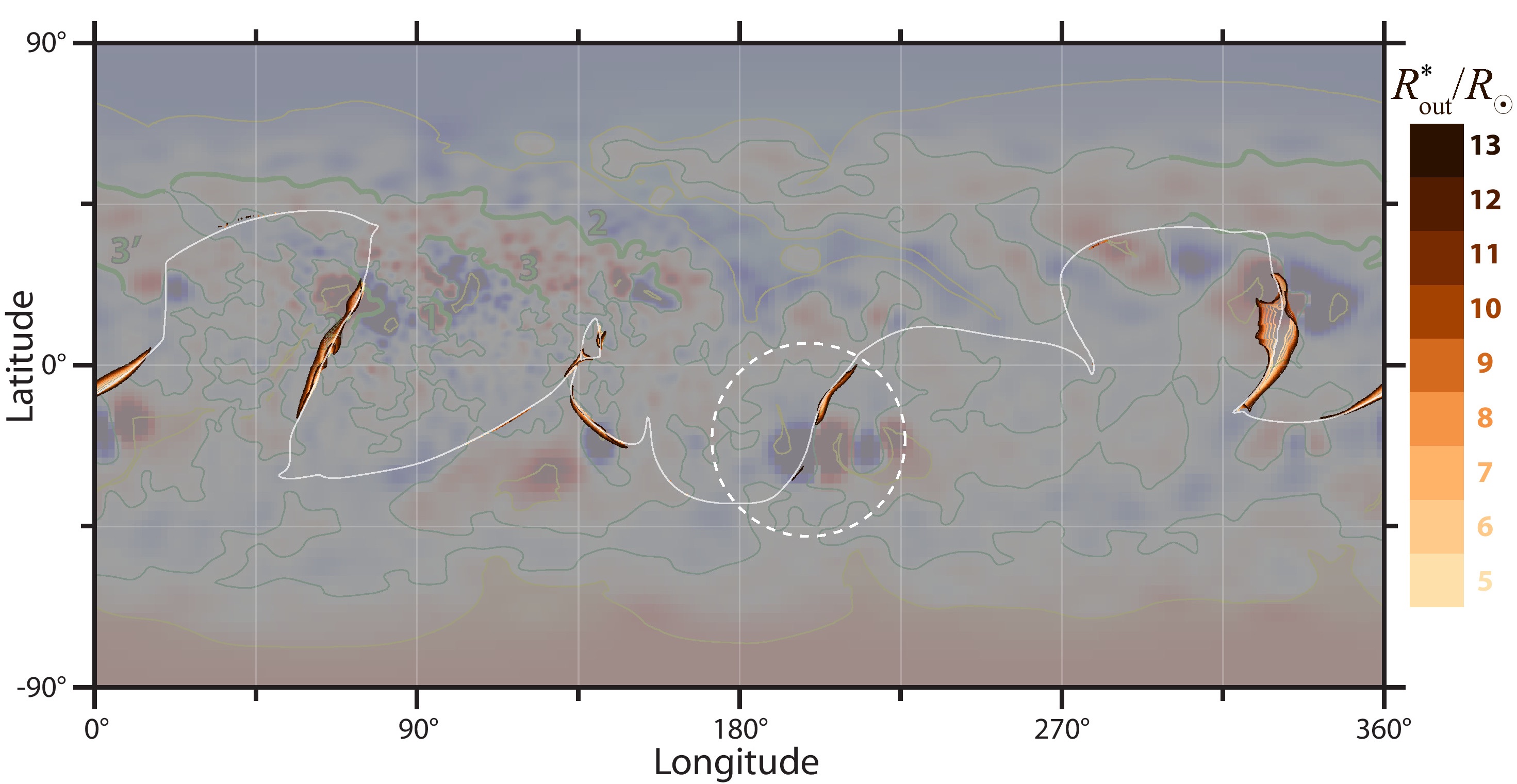}\\
%   \end{comment}
%\ \\
\caption
{
Maps of the contours enclosing disconnected-flux regions (DFRs): at fixed upper-boundary radius $\Rout^{*} \approx 5\, R_{\sun}$ and different times (top panel), and at fixed (final) time of the simulation and different $\Rout^{*}$ (bottom panel).
The time sequence covers the evolution of the DFRs from the instant they first appear in the domain $R_{\sun} \le r \le \Rout^{*}$. 
These maps are superimposed on the dimmed photospheric map of $B_{r}$ and coronal holes (Figure \ref{f:Br+CHs}).
The thin white line represents the PIL at $r=\Rout^{*}$, the dashed white line encircles a DFR that is transient for the chosen domain: it first appears at $r< \Rout^{*}$ and then moves to larger radii to become stabilized at $r\gtrsim 6\, R_{\sun}$.
}
	\label{f:DFregs_ev}
\end{figure*}

We have saved 15 instances of our relaxation run of the MHD model (Section \ref{s:model}) at equidistant times, starting from the initial potential field at instant 1.
The last nine configurations contain \fb{slowly evolving} DFRs, which are outlined in
the top panel of Figure \ref{f:DFregs_ev} by contours of different colors.
Once being formed in the HCS, the DFRs grow in size, but at a decaying rate,
except for one that behaves differently (encircled in Figure \ref{f:DFregs_ev}).
It has apparently a transient character by first emerging
and growing quickly for a while, and then\stth{ shrinking back into nonexistence within} \fb{disappearing from} the domain\stth{ under study}.
However, after inspecting how the DFRs vary with increasing $\Rout^{*}$ at instant 15 (bottom panel in Figure \ref{f:DFregs_ev}), we found that the transient DFR actually\stth{does exist in the computational domain, albeit} \fb{turns into a ``normal'' DFR located} at $r>\Rout^{*} \approx 5\, R_{\sun}$.
This means that an initially fast motion \fb{of the edge} of the \fb{transient} DFR\stth{ outward} \fb{away} from the Sun slows down at larger radii to a full stop.
Thus, all of the DFRs, including the transient one, eventually become\stth{ quasi-steady} \fb{almost stationary magnetic features with steady MHD flows in them}.

\fb{Scabbard-like} \gls{dfss} isolate the\fb{se} DFRs from surrounding open field \fb{(see Figure \ref{f:DFSSvsSC})}.
These surfaces are associated with the magnetic null points that reside in the HCS at a height of few solar radii,
either nearby or at the tips of the ``scabbards".
In the first case, they are formed by the field lines touching the upper boundary.
In the second case, they \stth{are} simply \fb{coincide with} a fan separatrix surface of one of the null points; this surface is folded and highly flattened inside the HCS.
In both cases, the field lines located at the opposite sides of the ``scabbard" show a strong shear,
which is due to \fb{a current-density} enhancement \stth{ of the}  \stth{enclosed} \fb{inside} the ``scabbard".
This is also the reason why the fans are folded and flattened at the related null points.

If the computational domain were extended to larger radii $\Rout$,
we would likely have at the end of the relaxation not several DFRs, but rather a single DFR covering the entire HCS.
The tendency toward such a merging of the DFRs can be inferred from their temporal and spatial variations presented in Figure \ref{f:DFregs_ev}. 
The top panel in particular shows that the quasi-steady DFRs, as defined for the domain with $\Rout^{*} \approx 5\, R_{\sun}$, grow in size and stretch with time, as if trying to merge with one another all over the HCS.
The bottom panel confirms this tendency in the radial variation of the DFRs at the final time, demonstrating that they indeed grow in size with increasing $\Rout^{*}$.
This implies that the gaps\stth{ among} \fb{between} the DFRs would shrink with increasing distance from the Sun to cause \fb{the} DFRs \stth{merging}\fb{to merge gradually} at larger radii.

In other words, our analysis suggests the following \stth{sketch of the global} magnetic topology \stth{in}\fb{of} the HCS \fb{in unbounded space}\stth{ with}\fb{, i.e., in the limit} $\Rout \rightarrow \infty$ and $\Rout^{*} \rightarrow \infty$.
\stth{In unbounded space, t}\fb{T}he \stth{helmet-streamer }field lines \fb{sucked into the HCS by the solar wind} cannot remain \fb{closed indefinitely}\stth{with growing radius}---due to stretching by the solar wind, their cusp summits will be sharpening \stth{inside the HCS} only up to certain critical heights.
Beyond these heights, the cusps of the\stth{ HCS} field lines have to flip over by switching their connectivity to the upper boundary, thereby forming a DFR.
The respective one-dimensional set of cusps at the critical \stth{radii}\fb{heights}, i.e., at the rim of this DFR, has to be no more than a multiple-null separator field line,
which zigzags and threads the null points residing at the tips of the ``scabbards".
We conjecture that such a global separator encircling the entire Sun \fb{in MHD models} is a substitute\stth{ in MHD models} for the source-surface null line in PFSS models.

The structure of the transient DFR, into which we are going to look in more detail now, is consistent with this \stth{sketch of the global magnetic topology}\fb{point of view}.
Transient DFRs are of particular interest, because they likely correspond to the observed disconnection events \citep[][]{DeForest2012} and so-called streamer blobs \citep[][]{Sheeley2009}.
To support this \stth{conjecture}\fb{statement}, we show in panel (a) of Figure \ref{f:s_blob} synthetic Thompson scattering brightness, as it would be observed in the solar north-south direction
 at time instant 11 in a coronagraph with the field-of-view covering $2-18\, R_\sun$.
An ad-hoc vignette function proportional to $(r-R_\sun)^3$ was used here to accentuate fainter structures at larger coronal heights.
The panel shows a compact region of enhanced brightness that is rapidly moving away from the Sun together with the lower edge of the transient DFR.
The angular width of this region is about $8^{\circ}$, which is comparable to the event width of $5^{\circ}$ estimated by \citet[][]{DeForest2012}.

The magnetic topology is rather intricate both inside this DFR and at its boundary (see panels (b) and (c), respectively).
The core of this structure is a magnetic ``roll" (panel (b)) consisting of fully disconnected field lines (yellow) that make many turns around a spiral null point before they return to the upper boundary.
The separatrix field lines (thick magenta lines) either spiral out from the null to the upper boundary, or go from it in opposite directions transversely to the spirals and down to the photosphere (see panel (b) and its inset).
Their footpoints \fb{lay} at the border of  the CH rooted in a strong positive polarity region (see the inset to panel (c)).
The field lines of the first type form a fan separatrix surface, while two field lines of the second type are spine separatrix field lines.
All these separatrix field lines are visualized here by using the BFLP technique described in Section \ref{ss:BFLP}.

%{\hR
%\citep{Gosling1995}
%}

Additional insights into the field line connectivity in this DFR are gained from combined upper-boundary maps of  $\slog Q$ and open/disconnected-flux regions, shown in panel (d).
The open-field regions on this map are colored in green.
To identify the flux tube footprints with a large field-line length $L$, the DFR is shaded in this map in purple if $L>100\, R_{\sun}$, and brown otherwise.
The purple footprints correspond to the magnetic blob, with field lines colored in yellow.
It is interesting that one of these footprints is not a simply-connected region:
It contains two narrow stripes of open-field regions, implying that two open-flux strands pierce through the volume of the transient DFR.
Several open field lines (white) of these strands are depicted in panels (b) and (c).
These field lines pass nearby the indicated spiral null point, which suggests that they are likely formed by reconnection at this null in the course of the blob moving away from the Sun.
 
The boundary magnetic surface of the transient DFR is a separatrix surface, which we recovered with the help of our BFLP technique.
It has four distinct parts, two of which belong to the open-field boundary and are rooted at the photosphere in positive and negative polarities.
These parts are distinguished in panel (c) of Figure \ref{f:s_blob} by
coloring the related BFLPs in red and blue, respectively, of various tints.
Three insets (ic1--ic3) to this panel show zoomed views of the BFLPs' branching at several null points.
The BFLPs abruptly split at them to give, in particular, V-type field lines that form the boundary of the transient DFR.
Sharp cusps of the V-type lines originating at the nulls indicate that the field-line structure in their vicinity is highly flattened, which reflects a local enhancement of the current density at this part of the DFR.
%latter is partly revealed in 
%These two parts join each other at relatively low heights 

With increasing number of used BFLPs, these null points tend to settle on a zigzag-like line segment.
Before splitting at the nulls, each of these BFLPs tangentially converges to a part of this segment, as seen from inset (ic3) to Figure \ref{f:s_blob} for a subset of the BFLPs.
This field-line structure suggests that the inferred segment is actually a multiple-null separator field line.
It is similar to the global separator described above, with one exception: It does not close on itself, as the global separator, but terminates at ends, which are two of the several nulls connected by the separator.
And it is exactly the separator along which the, oppositely directed, open fields reconnect with one another by replenishing adjacent closed and disconnected flux systems, below and above the separator, respectively.
Thus, with respect to the separator, the described open-field parts of the DFR boundary belong to the reconnection inflow region.

The remaining two parts of the DFR boundary are associated with the reconnection outflow region.
They consist of U-type field lines whose lower parts smoothly dip by curving near the end nulls of the separator. 
These disconnected field lines are separatrix lines as well, because one of their two footpoints is a contact point that belongs to the upper-boundary PIL.
In panel (c), they are continued from these contact points down to the photosphere as thinner lines of the same greenish colors.

\subsection{Magnetic Topology of Pseudo-Streamers}
	\label{s:ssPS}

As in the PFSS model, the pseudo-streamer lobes in our MHD configuration are delimited by the fan-like separarix surfaces that originate at either null points or bald patches \citep{Titov2011,Titov2012}.
These surfaces have the appearance of separatrix domes and the above-mentioned SCs.
Each SC is formed by both open and closed field lines, all connected to a single null point in the corona.
The null point is individual for each of the SCs and called basic null to distinguish it from other possible nulls related to the pseudo-streamer topology.
The separatrix surfaces are complemented by one-dimensional separatrix field lines \citep{Lau1990}, so-called spine lines \citep{Priest1996},  which are joined to the basic nulls transversely to these surfaces.
The angle between the line and the surface is the smaller the larger the local current density is \citep{Parnell1996}.

Figure \ref{f:DFSSvsSC} presents open-field segments of the three separatrix curtains (\SC{1}, \SC{2}, and \SC{3}), which presumably interacted with the erupting filaments 1--3, and 3' in the 2010 August 1--2 event.
For \SC{1} and \SC{3}, these segments are remarkably different to their PFSS analogues in the following respect:
When approaching the basic null, their BFLPs tend to osculate first, then make a sharp turn, and finally split at the null and go along the spine lines in opposite directions down to the photosphere.
In the PFSS model, the separatrix field lines of \SC{1} and \SC{3} instead approach their basic nulls directly, and they do so at different angles (see Figures 5 and 8 in Paper I), which implies some isotropy of the PFSS field in the fan planes near the nulls.

These structural differences in \SC{1} and \SC{3} between the models result from the electric currents that are induced in the MHD model in the vicinity of the basic nulls by solar-wind flows.
The latter have to be well aligned with the open field lines and so have to follow the highly curved paths of the field lines near the nulls.
This, in turn, implies the presence of significant inertia forces at these locations, which must be balanced in our \stth{quasi-steady }configuration by the Lorentz force and the pressure gradient.
Since the inertia forces are generally non-potential, the presence of the Lorentz force is necessary to sustain equilibrium. 
Thus, a concentration of the current nearby basic nulls, and hence a distortion of  their local field structures, are inevitable in MHD configurations.
It can be expected also that this effect becomes stronger with increasing height of the basic nulls, since the solar-wind speed in the low corona increases with height as well.
And indeed, we see this effect for \SC{1} and \SC{3}, but not for \SC{2}, whose basic null is at a much lower height than for the other two SCs.

There are also fans spanned by  closed field lines, which start either at other nulls or at BPs.
As in the PFSS model, each fan has the shape of a half-dome and delimits closed flux rooted in the parasitic polarity at one of the two flanks of a pseudo-streamer (see Figure 7 in Paper I).
The half-domes join each other along the spine line of the basic null to form the closed separatrix dome covering the parasitic polarity of a pseudo-streamer.
The whole structure is such that the SC intersects the dome along a separator field line and divides the enclosed flux of the parasitic polarity into two lobes, each of which may contain a filament (see Figures 5--7 in Paper I).

\subsubsection{Fine Structure of Pseudo-Streamer Separatix Curtains}  \label{s:fine_struc}

So far, we compared mainly the open-field parts of SCs in the pseudo-streamers present in our two magnetic field models.
By comparing also their closed-field parts we can gain even more interesting insights into the structure of SCs.
Figure \ref{f:SC1} illustrates this fact by showing three different views of \SC{1}.
Panel (a) presents the global view of \SC{1}, similar to the one used earlier in our PFSS model (cf. Figure 5 in Paper I).
Panel (b) shows a view of the other side of \SC{1}, and panel (c) depicts the same view, but\stth{ highly} zoomed into the area around the basic null of \SC{1}.
The open and closed separatrix field lines, colored here in two\stth{ slightly} different tints of cyan,
as well as thick pink field lines and thick magenta spine lines,
are all connected to the basic null.

The thick pink field lines in Figure \ref{f:SC1} touch the upper boundary and so represent the open separators.
As discussed above, their definition depends on how far away from the Sun this boundary is located.
However, the differences between the lower parts of the separators decrease when the upper boundary is moved farther out.
Moreover, since the upper ends of the pink field lines in this thought experiment remain inside the HCS, they will sooner or later join the far-distant nulls of the DFRs, which we discussed in Section \ref{s:DFRs}.
Once that is the case, we would have exact open separators connecting the basic null with the nulls of the DFRs.
Proving this conjecture is, however, technically challenging for a case study like ours,
because it would require a precise computation of the related very long BFLPs, which may, in principle, extend beyond $\Rout =20\,R_{\sun}$.
We postpone this task for the future, and focus here on the magnetic topology in our domain of radius $\Rout^{*} \approx 5 \, R_{\sun}$.

The closed-field parts of \SC{1}, which join its open-field part along the described separators, 
represent actually the transverse field-line structure of the helmet streamer and adjoint HCS at two distinct cross-sections.
A noticeable difference in the structure in these cross-sections is instructive, as it clearly demonstrates the presence of an inhomogeneous magnetic component that is aligned with the current of the HCS.
This property of the MHD-modeled configuration is already discussed in Section \ref{s:cmaps} in relation to the field lines forming the CHs' boundaries, to which the above open separators belong by definition.

We see in Figure \ref{f:SC1}a that the closed-field part of \SC{1} on the left
remains relatively wide almost all over its extension in height.
In contrast, the width of the closed-field part on the right rapidly shrinks with growing radius already from $r\simeq 2 R_{\sun}$ on.
This, however, does not necessarily mean that the HCS is wider on the left than on the right.
That is because \SC{1} crosses the HCS at these two sites at different angles: The crossing is nearly parallel to the HCS on the left and perpendicular to it on the right.
This, in turn, is because the magnetic component aligned with the current density is on average much larger in the left than in the right case.
%Thus, the variation of this component inside the HCS does explain the described difference in the appearance of the two closed-field parts of \SC{1}.

\subsubsection{Bifurcation of the Basic Null Point}  \label{s:bif_NP}

Of particular interest for our study is the region around the basic null, which is the point from which the two magenta thick lines emanate in Figure \ref{f:SC1}c, and it is the only null point present in the PFSS model in the same region.
This null survives the transition from the PFSS to the MHD model,\stth{ although it changes} \fb{but} 
its location and \stth{eigenvalue properties of its }\fb{matrix} of magnetic-field gradient\fb{s}\stth{ matrix} \fb{change}.
Moreover, as the panel shows, the basic null undergoes also a bifurcation during this transition, giving birth to two radial and two spiral nulls that are indicated in Figure \ref{f:SC1}c by white and black arrows, respectively.

Physically, this bifurcation is likely due to the formation of plasmoids by the tearing instability in a current layer that is formed near the basic null due to MHD flows.
Each plasmoid has a spiral null point at its center, with its spines co-aligned locally with the current density.
A similar bifurcation, but into one spiral and two radial nulls, was described by  \citet{Wyper2014b, Wyper2014a}  for jet-type configurations with a simpler initial spine-fan topology.
They explained it by analogy with a two-dimensional bifurcation of an X-point into one O- and two X-points such that ${\rm X \rightarrow X-O-X}$.
Accordingly, we can say that our case corresponds to the bifurcation of an X-point into two O- and three X-points as follows: ${\rm X \rightarrow X-O-X-O-X}$.
The new X-point that corresponds to our basic null would be located between the O-points, which correspond in three dimensions to our spiral nulls.

To distinguish individual field-line structures associated with different nulls, the corresponding BFLPs are colored in Figure \ref{f:SC1} differently.
In particular, the BFLPs of the spiral nulls are depicted by orange tubes,
while the BFLPs of the radial null at the lowest height are represented by thinner yellow tubes.

The cyan tubes depict both open and closed BFLPs of the basic null, which is located exactly in between the spiral nulls.
These BFLPs descend from the upper part of the structure, rapidly converge toward one another, and form a collimated bunch of field lines (see Figure \ref{f:SC1}c).
This bunch then makes a sharp turn and starts skirting the upper spiral null toward the basic null. 
The cyan BFLPs from the lower part of the structure also approach this null, but at different angles.
Reaching the basic null, each pair of the cyan BFLPs abruptly splits and forms two new bunches that emanate from the null in opposite directions along the spine lines, which are represented by thick magenta tubes\stth{ in the figure}.
The new bunches are not visible \fb{in the figure},\stth{ since} \fb{because} they are\stth{thinner than the spines} \fb{enclosed by the thicker magenta tubes}.

Figure \ref{f:SC1}c does not show the BFLPs of the third radial null, but it shows its location (the highest one among all shown nulls),
which is where the cyan BFLPs of the basic null abruptly turn after their rapid convergence.
The turning point is slightly off from the third radial null, which allows these BFLPs to be deflected in one direction, i.e., to continue without splitting.
The (not-depicted) BFLPs of the third null closely follow the cyan BFLPs and then split at the null into two bunches, as expected.
One of them merges with the bunch of the cyan BFLPs, while the other first moves in the opposite direction and then runs parallel to the magenta spine toward the northern polar CH.
Thus, these two bunches approximate the spines of the third radial null,
while the fan surface of that null nearly coincides with that of the basic null.
In other words, the separatrix curtain \SC{1} in the MHD model actually consists of two very close fan surfaces, one of which originates at the basic null, while the other at the third radial null;
both these surfaces are embedded into a thin QSL.

This consideration helps\stth{ one} to understand how \SC{1} is topologically linked to the northern polar CH.
In the top panel of Figure 2, the latter appears to be extending as a high-$Q$ line toward the location of filament 2.
Refining the field-line mapping, we find that this line indeed contains a tendril-like outgrowth of the CH, with its width narrowing down to zero approximately at the midpoint of that line.
This point is at the very tip of the open-field tendril, and it serves as a footpoint for the spine line that links to the third radial null.
Thus, this spine is an edge of the volumetric wedge-like outgrowth of the northern polar CH.
Its other two edges nearly coincide with the open separators (thick pink lines in Figure \ref{f:SC1}).
The small differences between them are due to the minor offset of the third radial null from the turning point of the cyan BFLPs.

The remaining part of the high-$Q$ line -- the one that is closer than the other to the location of filament 2 --
is due to closed field lines that run in the vicinity of a small isolated negative polarity (see the top panel in Figure 2).
The latter is embedded into the positive parasitic polarity of the pseudo-streamer and, hence, can be regarded as a ``secondary" parasitic polarity.
It is encircled by the PIL with a BP at its northern side, so that
the field lines touching the BP form a dome-like surface that separates the isolated polarity flux from the ambient one in a manner similar as described by \citet{Bungey1996}.

However, near the top of this dome, there exists a magnetic minimum point where the field drops by approximately three orders of magnitude compared to its value averaged over the region under the dome.
This magnetic minimum gives birth to a hyperbolic flux tube whose one footprint is exactly the discussed closed-field part of the high-$Q$ line.
The thin gray tubes in Figure \ref{f:SC1}c present the corresponding BFLPs, which reveal the typical structure of a hyperbolic flux tube originating at a magnetic minimum.
It is characterized by a rapid convergence of the field lines in one transverse direction and their rapid divergence in the other \citep{Titov2007a, Titov2009}.
These two directions correspond, respectively, to the vertical and horizontal quasi-separatrix surfaces visualized in the figure by the gray BFLPs.
The hyperbolic flux tube provides a continuous transition of the structure
from the BP separatrix dome of the ``secondary" parasitic polarity to the upper-lying \SC{1} and the adjacent volumetric outgrowth of the northern polar CH.
Altogether, this forms a complex set of (quasi-)separatrix surfaces and lines that accommodates the reconnection between closed and open field lines in the pseudo-streamer.
Possible implications of this process for the dynamics of the coronal plasma will be discussed in Section \ref{s:impl}.

It is instructive for comprehending the considered multi-null magnetic topology to pay attention in Figure \ref{f:SC1}c to the field direction at the BFLPs, as well as to their alignments with other structural elements of the pseudo-streamer.  
In particular, at the spiral nulls, the magnetic field at their spines and fans is directed toward and outward of the null, respectively.
The spines are parallel to \SC{1}, while the fans are aligned with the quasi-separatrix surface adjacent to the outgrowth of the northern polar CH.
At the radial nulls, the respective directions and alignments of the spines and fans are opposite.

In conclusion, we emphasize that the complexity of the magnetic topology is revealed here by applying just our BFLP technique.
All topological features are immediately recovered by this technique as soon as the related high-$Q$ lines are populated with BFLP footpoints densely enough to have all nulls in the region of interest bracketed.

\subsubsection{Comments on Separatrix Domes}

Comparing the photospheric $\slog Q$-maps in Figure \ref{f:slogQ_maps} (top panel) and in Figure 3 in Paper I for closed field regions, we see that they are very similar in the MHD and PFSS configurations.
This implies that the related separatrix domes are similar in both models as well.
Therefore, Figure 7 in Paper I provides a good proxy for the separatrix domes that the MHD model yields in the source region of the first three eruptions on 1-2 August, 2010.
The half-domes that join each other along the spines of the null ${\rm N}_{1}$ and cover flux ropes 2 and 3 are basically due to two modifications in the MHD model.
First, they are slightly pulled upward by the solar wind.
Second, they form a cusp structure at their junction line along the spines of the basic null by matching in their cross-section the osculation that the open separators have at this null (see Figure \ref{f:SC1}).

As in the PFSS model, the half-domes intersect\stth{, of course,} the separatrix curtain \SC{1} along the closed separator field lines, which are all connected at one end to the basic null point.
They are not shown explicitly in Figure \ref{f:SC1} for simplicity, but their loop-like shapes can easily be inferred from\stth{ viewing} \fb{considering} the closed field lines of \SC{1}.
\fb{Similar closed separators are depicted in Figures 5--8 of Paper I as thick red tubes.}

One should be cautious, though, about the good correspondence between our two models for closed separatrix structures in the low corona,
because it is mainly a result of the used photospheric boundary conditions for the magnetic field.
Only the radial field component derived from the synoptic map is used in both models to calculate the configurations,
which leads to a nearly potential field at low heights in the MHD model.
We anticipate, however, that MHD magnetic-field structures reconstructed from vector magnetic data would significantly differ from their PFSS analogues.
Therefore, we refrain here from a more detailed comparison of the separatrix domes of our two models. 

\section{SOME SPIN-OFF IMPLICATIONS OF OUR ANALYSIS}
	\label{s:impl}

As suggested by our analysis, pseudo-streamers have several model-independent properties that likely support coupling between successive CMEs.
Two of them are also interesting in connection to the mechanisms of particle acceleration in solar eruptions:
(1) The parasitic-polarity PIL in pseudo-streamers appears to be a favorable site for the formation of erupting flux ropes;
(2) A pseudo-streamer has always a separatrix curtain that starts at a basic null point and contains two open and several closed separator field lines, all connected to that null.

These separators \fb{lay} at the intersections of the separatrix curtain with other separatrix surfaces such as the closed half-domes and the helmet-streamer dome, which are connected to one another along the spine lines of the basic null point.
The separatrix surfaces divide the volume around the parasitic polarity of the pseudo-streamer into topologically distinct cells of magnetic flux.
Flux ropes can reside initially in the two lowest cells, i.e., in the closed-flux lobes delimited by the separatrix curtain and half-domes.
As demonstrated in our idealized MHD simulation \citep{Torok2011}, an eruption of one of the ropes has to cause a significant flux redistribution between that cells by triggering magnetic reconnection along the separators, which is one of the suggested mechanisms for the acceleration of particles \citep[][]{Priest2000}.
Since the reconnection proceeds generally along the entire length of separators \citep{Parnell2010} and, in our case, in the low corona where the magnetic and electric fields should be relatively strong, the related acceleration of particles should be rather efficient.
What is also critically important here is that the separators, as well as the open field lines of the separatrix curtain, are all connected to the basic null point, which
facilitates  the escape of the accelerated particles into the open corona.
Thus, pseudo-streamers appear to be uniquely favorable sites for both the acceleration of solar energetic particles (SEPs) and their injection into interplanetary space.

If this conjecture is true, the existent simple picture of SEP event classification  \citep[see][]{Reames2013} becomes blurred.
According to this classification, there are only two distinct types of SEP events, namely, impulsive and gradual ones.
Supposedly, each of them has a distinct underlying mechanism of particle acceleration and a distinct supporting magnetic topology \citep[][]{Reames2002}.

In impulsive SEP events, the charged particles are accelerated in flares by resonant wave-particle interactions in a turbulent reconnection region.
The latter is assumed to develop between interacting closed and open fields of locally opposite orientations. 
The simplest 3D configuration where open and closed fields contact each other is one with a fan-spine magnetic topology \citep{Masson2009}.

In gradual SEP events, the acceleration of particles occurs at the shock wave that is formed ahead of the erupting flux rope in the corona and heliosphere \citep[e.g.,][]{Schwadron2015}.
Those particles that are accelerated earlier in the reconnection region beneath the flux rope cannot escape along open magnetic field lines and so have to precipitate down to the chromosphere. 
However, this commonly accepted two-class picture of SEP events is challenged by observations of events with\stth{ energetic particle} \fb{SEP} populations\stth{ that show} \fb{whose} abundance ratios\stth{ that} are consistent with both flare and shock origin \citep[e.g.,][and references therein]{Cane2006, Desai2016}.
%However, this commonly accepted two-class picture of SEP events is challenged by observations of unusual large events, which were presumably ``gradual" SEP events but looked ``impulsive" at energies $>10\:{\rm MeV}$ \citep{Cliver2009, Cliver2016}.

In this regard, we note that the magnetic topology of pseudo-streamers implies the existence of new type SEP events that combines these two basic types.
Indeed, as explained above, the eruption of the flux rope, residing initially in one of the pseudo-streamer lobes, can accelerate energetic particles along the separators connected to the basic null point.
The open part of the separatrix field lines, also connected to the null, allows accelerated particles to escape,
in a similar fashion as in impulsive SEP events.

However, if the eruption of the flux rope is fast enough, it has also to lead to the formation of a shock wave and associated particle acceleration, typical for the gradual SEP events.
This implies that eruptions that originate in pseudo-streamers may
produce two successive SEP events---first impulsive, then gradual. 
(This timing might be difficult to distinguish at 1 AU from in situ measurements, and very dependent on the field lines sampled by the spacecraft.) 
Our case study suggests that, from the topological point of view, this scenario may occur quite often, so its further investigation
is of significant interest.

This new scenario for SEP events is similar to the scenario that \citet{Masson2013} have recently proposed for an eruptive configuration containing both open and closed magnetic fields.
The closed field has initially a sheared magnetic arcade formed inside a dome-like fan separatrix surface that originates at a coronal null point with locally vertical inner and outer spine lines.
The authors demonstrated numerically that the eruption of the sheared arcade leads to the formation of an erupting flux rope and two reconnecting current sheets---one of them below the rope and the other at the null point.
They showed that if this process takes place close to an open-field boundary, it will cause reconnection between open and closed fields in the current sheets, which naturally leads to the escape of accelerated particles into the open field.

The \fb{discussed} scenario for SEP events is not the only spin-off implication of our analysis of pseudo-streamer magnetic topology. 
Another potentially important implication arises in connection to the slow solar wind problem. 
Upgrading our PFSS configuration to MHD model, we found so-called ``plasmoids'', which have a peculiar 3D magnetic structure.
Each of them consists of two flux ropes of opposite handedness, meaning that their axial currents are co-directed, while their axial fields are counter-directed.
These ropes are rooted with one end in the photospheric regions of the like polarity and, with their other ends, are conjoined with each other at a spiral null point.
Such a system of two conjoined flux ropes is a 3D analog of plasmoids produced in reconnecting current layers due to the tearing instability in two-dimensional plasmas \citep[][]{Priest2000}
and was described by  \citet{Wyper2014b, Wyper2014a}  for configurations with a spine-fan magnetic topology.
Our analysis suggests that they appear at least in one of our pseudo-streamers near the basic null of separatrix curtain  \SC{1}, where a current layer is formed and bifurcated into several plasmoids.
The ongoing interchange reconnection in this current layer provides transfer of the open flux between \CH{1} and the northern polar CH, likely as a residual part of the MHD relaxation to a full\stth{y steady configuration} \fb{equilibrium}.

The dynamics of the plasmoids in the current layer should cause some pulsations in its neighborhood, which must freely propagate along the adjacent open field lines of the separatrix curtain outward in the corona.
The details of this process have yet to be properly studied, but based on the already gained insights, we speculate that such pulsations might naturally explain an enhanced variability of the slow component of the solar wind.
This is because the same process probably occurs at other separatrix curtains that numerously surround the HCS  and where, according to the S-web model \citep{Antiochos2011, Linker2011, Titov2011}, the slow solar wind is formed.

In addition, a similar plasmoid, also consisting of two flux ropes conjoined at a spiral null point, appears in our investigation at a much larger scale of the size of several solar radii.
As discussed in Section \ref{s:DFRs}, the transient DFR created during the MHD relaxation has also a spiral null at the center of an erupting magnetic blob, which moves ahead of a streamer plasma blob.
Note that the spine lines of the null are rooted at the same photospheric polarity (see panels (b) and (ic1) in Figure \ref{f:s_blob})) and its neighboring field-line structure locally resembles the plasmoids of the pseudo-streamer.
This suggests, on the one hand, that the magnetic blob is nothing else than a huge plasmoid\stth{of} \fb{within} the HCS.
Propagating radially outward from the Sun, it transports both magnetic and mass fluxes into the upper corona and heliosphere.
On the other hand, one could speculate that \stth{the}\fb{such} plasmoids play a similar dual role in pseudo-streamers, but on much smaller length scales, more difficult to observe.
%Very recently, conjoined flux ropes have been found also in flux-emergence simulations (T\"or\"ok et al., in preparation), so these structures may be quite common in the corona.

%%%%%%%%%%%%%%%%%%%%%%%%%%%%%%%%%%%%%%%%%%%%%%%%%
%%
%%                      Summary
%%
%%%%%%%%%%%%%%%%%%%%%%%%%%%%%%%%%%%%%%%%%%%%%%%%%

\section{SUMMARY AND DISCUSSION}
	\label{s:s}

In this work, we analyzed the global topology of the background magnetic field for a sequence of CMEs that occurred on August 1--2, 2010 \citep{Schrijver2011, Torok2011}.
The background field was reconstructed from photospheric magnetograms by using our MHD model that incorporates plasma thermodynamics.
Previously, we have made a similar analysis for a PFSS model derived from the same data \citep{Titov2012}. 
As for the PFSS model, we find that each of the three source regions of the CMEs contains a pseudo-streamer that separates neighboring coronal holes of negative polarities from one another.
The MHD model confirms that each of these pseudo-streamers starts at the photosphere as a positive parasitic polarity that is divided in the middle by a vertical magnetic surface, called separatrix curtain. 
The field lines of the separatrix curtain, both open and closed, fan out from a so-called basic null point.
Among them are two open separator field lines that \fb{lay} at the intersections of the separatrix curtain with the helmet-streamer separatrix dome.
Therefore, the open separators are the only sites where the open and closed magnetic fields contact each other.
Thus, they are the very sites where the so-called interchange reconnection between closed and open fields should take place as the fields evolve.

There are also separatrix half-domes that are joined together on the spine lines of the basic null to cover the parasitic polarity.
These domes are intersected by the separatrix curtain along the closed separator field lines, which are all connected at one end to the basic null point.
Reaching the photosphere, the separatrix curtain divides the closed field of the parasitic polarity into two lobes, each of which may harbor a magnetic flux rope.
This topological structure implies that by producing a chain of sympathetic eruptions,\stth{those} \fb{such} flux ropes have to cause magnetic reconnection along those separators, to provide the necessary redistribution of the flux in the configuration.
In this way, the topological structure sets a coupling between\stth{ individual} \fb{sequential} eruptions, which was the essence of our idealized MHD simulation of sympathetic eruptions \citep{Torok2011}.
Thus, the present work further substantiates that model and reasserts its main conclusion on the key role of the pseudo-streamer magnetic topology in linking such eruptions.

We have also identified salient differences between magnetic structures in our MHD and PFSS models.
Those are the \fb{shapes,} sizes\fb{,} and flux contents of the coronal holes, which are significantly larger for the MHD model, particularly in the source regions\stth{, as well as the shapes of the coronal holes}.
These differences become even more prominent in the high corona, as clearly revealed in our $Q$-maps and coronal holes' connectivity maps.
They manifest in rather different shapes of the HCS and coronal holes' boundaries, outlined by high-$Q$ arcs, and in the different ways the coronal hole boundary fluxes split and connect in the HCS to one another.
Thus, our case study calls for reassessing the accuracy of the PFSS approximation in reproducing the global structure of the coronal magnetic field near the HCS.
Based on the obtained results, we have to state firmly that its MHD description cannot be matched satisfactorily by a PFSS model.

The MHD model additionally provides a number of less obvious but still important insights into the large-scale topology and reconnection of the magnetic field in the corona.
One of them is the generalization of the open separator field line, defined now as a field line lying at the intersection of a separatrix curtain and a helmet-streamer separatrix dome.
The properties of this dome noticeably depend on the model used for producing the global configuration.
In a PFSS model, the dome originates at the source-surface null line, which is unfortunately not a stable topological feature---it has to disappear when switching to a more realistic magnetic field model, such as the one provided by MHD.
Our analysis suggests that in this case, instead of the null line, there should be a multiple-null separator field line\stth{ lying} \fb{located} at the very top of the helmet-streamer separatrix dome, well inside the HCS.
Some of the nulls of this global separator have to be also upper end points of the open separators of pseudo-streamers.
In addition,\stth{ such a} \fb{this} global separator is the edge of disconnected flux regions that extend the HCS far into the heliosphere and interplanetary space.
Thus, this is the line along which the closed, disconnected, and open flux regions come into contact, and where reconnection between them occurs to support the global evolution of the coronal magnetic field.

However, the picture just described refers to an ideal situation where one would be able to model the global magnetic field and analyze its topology for a domain extending over many solar radii.
For smaller domains like ours, the helmet-streamer separatrix dome differs from the one discussed above.
It is approximated by the surface that is mainly formed by the field lines touching the upper boundary at its polarity inversion line, while the other parts of the surface---the separatrix boundaries of the disconnected flux regions---are only partially included in it, as they are formed at larger radii outside the analyzed volume.
For this reason, the above description of the global magnetic topology is only inferred rather than determined; more technical efforts are required to prove this conjecture explicitly by extending the analysis to a larger volume.

In fact, the MHD flows bring even more complexity into the topology of the configuration.
For example, they force the basic null of the pseudo-streamer to bifurcate into several other nulls, all located in the current layer near the basic null.
Some of them are spiral nulls centered in the \stth{midst}\fb{middle} of 3D plasmoids, which are basically two flux ropes of opposite handedness conjoined at the fan separatrix surface of the associated spiral null.
These plasmoids are likely produced by the tearing instability developing in the current layer, and have to play a certain role in the transfer of mass and magnetic fluxes into the solar wind.

A similar plasmoid, but on much larger length scales, is present in the transient disconnected flux region of the HCS during its relaxation to a\fb{n} \stth{quasi-steady state}\fb{equilibrium}.
A fast upward retraction of its U-shaped magnetic loops scoops up coronal plasma by forming a dense blob in the lower part of the plasmoid, both of which rapidly propagate outward into the corona.
This process is accompanied by a reconfiguration of the global magnetic field via reconnection of open magnetic fields below the blob\stth{ very} low in the corona.
It appears that this intriguing phenomenon corresponds to streamer blobs in observed disconnection events and so deserves a more detailed investigation in the future. 

Finally, our analysis of the pseudo-streamer magnetic topology also reveals interesting implications for understanding SEP acceleration in CMEs, driven by erupting flux ropes that resided initially in pseudo-streamer lobes.
In particular, it predicts that such eruptions could sequentially produce a pair of related impulsive and gradual SEP events, which so far have been considered as two distinct types of events.

%%%%%%%%%%%%%%%%%%%%%%%%%%%%%%%%%%%%%%%%%%%%%%%%%
%%
%%                           Acknowledgements
%%
%%%%%%%%%%%%%%%%%%%%%%%%%%%%%%%%%%%%%%%%%%%%%%%%%

\acknowledgments

The contribution of V.S.T., Z.M., T.T., and J.A.L. was supported by NASA's HSR, LWS, and HGI programs,
NSF Grants AGS-1156119 (SHINE program) and AGS-1560411, NSF (Frontiers in Earth System Dynamics program), and contracts from Lockheed-Martin and Stanford University to Predictive Science, Inc.; O.P. was supported by NASA Grant NNX15AB89G.
Computational resources for the MHD computation were provided by NASA Advanced Supercomputing and the NSF supported XSEDE program.
%%%%%%%%%%%%%%%%%%%%%%%%%%%%%%%%%%%%%%%%%%%%%%%%%
%%
%%                           Appendix
%%
%%%%%%%%%%%%%%%%%%%%%%%%%%%%%%%%%%%%%%%%%%%%%%%%%

\appendix

\section{BALD-PATCH CRITERION FOR A CURVED BOUNDARY}
\label{s:bpc}

The criterion for the presence of the BP at a given PIL was earlier derived by \citet[][]{Titov1993}\stth{ under the assumption that} \fb{for}\stth{the} \fb{a flat} boundary\stth{ surface is flat}.
Therefore, its generalization for our spherical boundary is required.
In fact, such a generalized criterion is applicable to an arbitrary \stth{smooth}\fb{curved} surface as well, since the latter can locally be approximated by a sphere.

Let us derive this criterion for a sphere of radius $R_{\sun}$ at which the radial component of the magnetic tension force in spherical coordinates is proportional to
\begin{eqnarray*}
\left.\left[ \hat{\bf r}\cdot\left({\bf B \cdot \nabla}\right){\bf B} \right] \right|_{r=R_{\sun}}=
  \left. \left({\bf B \cdot \nabla} B_{r} \right) \right|_{r=R_{\sun}}  - \frac{\left.\left( B_{\theta}^{2}+B_{\phi}^{2}\right)  \right|_{r=R_{\sun}} }{R_{\sun}} \, ,
\end{eqnarray*}
where $\hat{\bf r}$ is a unit radial vector.
As $B_{r}=0$ at the PIL, ${\bf B}$ coincides there with the tangential field ${\bf B_{\tau}}$, so that
\begin{eqnarray}
\left.\left[ \hat{\bf r}\cdot\left({\bf B \cdot \nabla}\right){\bf B} \right] \right|_{\rm PIL}=
  \left. \left({\bf B_{\tau} \cdot \nabla} B_{r} \right) \right|_{\rm PIL}  - \frac{\left.\left( B_{\theta}^{2}+B_{\phi}^{2}\right)  \right|_{\rm PIL} }{R_{\sun}} \, .
	\label{MTr1}
\end{eqnarray}

However, we can also write, in general, 
\begin{eqnarray*}
 \left({\bf B \cdot \nabla} \right) {\bf B} = \frac{{\bf B}^{2}}{\cal R} \, \hat{\bf n}  \, ,
\end{eqnarray*}
where ${\cal R}$ is a local curvature radius of the field line and $\hat{\bf n}$ is a unit normal vector to it.
From this formula, we obtain the expression alternative to (\ref{MTr1}), namely,
\begin{eqnarray}
\left.\left[ \hat{\bf r}\cdot\left({\bf B \cdot \nabla}\right){\bf B} \right] \right|_{\rm PIL} = \left. \left[ \frac{B_{\theta}^{2}+B_{\phi}^{2} }{\cal R} \, (\hat{\bf r} \cdot \hat{\bf n}) \right] \right|_{\rm PIL} 
= \left. \left[ \frac{B_{\theta}^{2}+B_{\phi}^{2} }{R_{\sun} } \, \frac{R}{{\cal R} }  \, \sign(\hat{\bf r} \cdot \hat{\bf n})\right] \right|_{\rm PIL}\, ,
	\label{MTr2}
\end{eqnarray}
where $R= R_{\sun}\, |(\hat{\bf r} \cdot \hat{\bf n})| \le R_{\sun} $ is the radius of the circle of intersection of our sphere with the osculating plane spanned on the local vectors ${\bf B}$ and $\hat{\bf n}$.
Combining equations (\ref{MTr1}) and (\ref{MTr2}), we obtain
\begin{eqnarray}
  \left. \left({\bf B_{\tau} \cdot \nabla} B_{r} \right) \right|_{\rm PIL} = \left. \left[ \frac{B_{\theta}^{2}+B_{\phi}^{2} }{R_{\sun} } \, \left( \frac{R}{{\cal R} }\, \sign(\hat{\bf r} \cdot \hat{\bf n})  + 1\right) \right] \right|_{\rm PIL}
  \, .
\end{eqnarray}
This expression is positive for concave field lines touching the sphere at the PIL, as $(\hat{\bf r} \cdot \hat{\bf n})$ is positive for them. 
For convex field lines, however, it is positive iff ${\cal R}>R$, which is exactly what is required for \stth{the}\fb{a} touching field line to be locally above the boundary at the contact point.
Thus, the inequality
\begin{eqnarray}
  \left. \left({\bf B_{\tau} \cdot \nabla} B_{r} \right) \right|_{\rm PIL} > 0 \, 
	\label{bpc}
\end{eqnarray}
is a necessary and sufficient condition for the BP presence.
It is independent of the curvature of the boundary surface, although its derivation assumes that the normal $\hat{\bf r}$ is to be directed oppositely to the curvature radius of the surface.
As in the case of \stth{the}\fb{a} plane boundary \citep[][]{Titov1993}, this criterion simply requires that {\it the magnetic field at the BP has to \stth{have the anomalous direction,}\fb{be directed} from negative to positive magnetic polarity}.

Being applied to the upper boundary, equation (\ref{bpc}) determines those segments of the PIL which belong to the disconnected flux regions.
The remaining segments of the PIL represent the apex of the helmet-streamer separatrix dome.

\section{HELMET-STREAMER SEPARATRIX DOME NEAR THE OUTER BOUNDARY}
\label{s:HSSDob}
%%%%%%%%%%%%%%%%%%%%%%%%%%%%%%%
%%%%%%%%  see calculations in  BP_HCS_expansion.mw
%%%%%%%%%%%%%%%%%%%%%%%%%%%%%%%

The upper-boundary PIL is determined by the equation $B_{r}(\Rout^{*},\theta, \phi) = 0$.
Any function restricted on this line will hereafter be designated by subscript ${\rm PIL}^{\!\!*}$.
Depending on whether we deal with a PFSS or MHD configuration, \fb{either} the entire PIL or only its subset, respectively, serves as an apex for the helmet-streamer separatrix dome.
For an MHD configuration, by launching field lines from the apex one can retrieve the dome itself.
%, as its magnetic field does not vanish at generic apex points.
In a PFSS configuration, however, the apex is a null line, so its points cannot directly be used to start field-line integration.
For these purposes, one should use, instead, appropriate neighboring points.
Assuming below that $\Delta l$ is a small distance between them and the apex points, we derive their locations in terms of small corrections $(\Delta r, \Delta\theta, \Delta\phi)$ to the apex-point coordinates.
This provides the structure of the helmet-streamer separatrix dome near its apex.
For the purpose of comparison, we derive it for both types of configurations, starting from the simpler MHD type.
%For the sake of comparison and completeness, we do it first for MHD configurations.

\subsection{MHD configuration}

In this case, the magnetic field at the upper-boundary PIL
\begin{eqnarray}
   B_{{\rm PIL}^{\!\!*}} \equiv\left. \left( B_{\theta}^{2} + B_{\phi}^{2} \right)^{1/2} \right|_{{\rm PIL}^{\!\!*}}
\end{eqnarray}
generally does not vanish, so one can seek solution of the field-line equation in the form of Taylor series in $\Delta l$.
The leading terms of this series yield
%
%\begin{eqnarray}
%        & \quad &  {\rm MHD}  \quad  {\rm PFSS}
%        \nonumber \\
%	\label{DrMHD}
%  \Delta r         & = & \frac{\left. \left( {\bf  B} \cdot \nabla B_{r} \right) \right|_{{\rm PIL}^{\!\!*}}}{B_{{\rm PIL}^{\!\!*}}^{2}} 
%\, \Delta l^{2}  
%                        ,   - \frac{\Delta l }{\sqrt{2}}  \, , \\
%	\label{DtMHD}
% \Rout^{*}\, \Delta \theta & = & \pm\left. \frac{ B_{\theta}}{B} \right|_{{\rm PIL}^{\!\!*}}  \Delta l 
%                                        ,  \mp  \frac{\left. \displaystyle \frac{\partial B_{r}}{\partial \theta} \right|_{{\rm PIL}^{\!\!*}} \Delta l  }
%                     { \sqrt{2}\, \Rout^{*} \left| \nabla B_{r} \right|_{{\rm PIL}^{\!\!*}} }  \, , \\
%	\label{DpMHD}
% \Rout^{*} \sin \theta_{{\rm PIL}^{\!\!*}}\, \Delta \phi    & = & \pm\left. \frac{ B_{\phi}}{B} \right|_{{\rm PIL}^{\!\!*}}  \Delta l 
%                                           ,  \mp \frac{ \left. \displaystyle \frac{1}{\sin\theta} \frac{\partial B_{r}}{\partial \phi} \right|_{{\rm PIL}^{\!\!*}} \Delta l  }
%                   { \sqrt{2}\, \Rout^{*} \left| \nabla B_{r} \right|_{{\rm PIL}^{\!\!*}} }  \, ,
%\end{eqnarray}
%
\begin{eqnarray}
	\label{DrMHD}
  \Delta r         & = & \frac{\left. \left( {\bf  B}_{\tau}  \cdot \nabla B_{r} \right) \right|_{{\rm PIL}^{\!\!*}}}{B_{{\rm PIL}^{\!\!*}}^{2}} \, \Delta l^{2}  \, , \\
	\label{DtMHD}
 \Rout^{*}\, \Delta \theta & = & \pm\left. \frac{ B_{\theta}}{B} \right|_{{\rm PIL}^{\!\!*}}  \Delta l \, , \\
	\label{DpMHD}
 \Rout^{*} \sin \theta_{{\rm PIL}^{\!\!*}}\, \Delta \phi    & = & \pm\left. \frac{ B_{\phi}}{B} \right|_{{\rm PIL}^{\!\!*}}  \Delta l \, .
\end{eqnarray}
Equation (\ref{DrMHD}) implies that $\Delta r < 0$ iff 
\begin{eqnarray}
      \left. \left({\bf B_{\tau} \cdot \nabla} B_{r} \right) \right|_{{\rm PIL}^{\!\!*} }
    \equiv  \left. \left(    \frac{B_{\theta}} {r}  \frac{\partial B_{r}} {\partial \theta}
                           + \frac{B_{\phi}} {r\sin\theta}  \frac{\partial B_{r}} {\partial \phi} 
        \right) \right|_{{\rm PIL}^{\!\!*} }
    < 0 \, ,
\end{eqnarray}
which is opposite to inequality (\ref{bpc}), as required.
The upper and lower signs in equations (\ref{DtMHD}) and (\ref{DpMHD}) correspond to backward and forward, respectively, displacements from the apex relative to the local magnetic field.

\subsection{PFSS configuration}
% \left| {\bf  B} \cdot \nabla B_{r} \right|
The PFSS boundary condition
$
  \left. {\bf B}_{\tau} \right|_{r=\Rout^{*}} = 0
	\label{PFSSbc}
$
implies that the matrix of tangential gradients of ${\bf B}_{\tau}$ is vanishing, i.e.,
$
  \left. \left[ \nabla_{\tau} {\bf B}_{\tau} \right] \right|_{r=\Rout^{*}}  = 0
$
. Taking also into account that the magnetic field is potential, i.e., ${\bf B} = - \nabla F$ such that $\nabla^{2} F =0$ holds everywhere up to the boundaries, one can obtain that
$
  \left. \frac{\partial B_{r}}{\partial r} \right|_{{\rm PIL}^{\!\!*}} = 0
$
and
\begin{eqnarray}
    \left. \left[  \nabla {\bf B}\right]\right|_{{\rm PIL}^{\!\!*}} =
    \left. \left[  \nabla {\bf B}\right]^{\rm T}\right|_{{\rm PIL}^{\!\!*}} =
    \frac{1}{\Rout^{*}}
\left.\left(
\begin{array}{ccc}
  0                                                                                   &\frac{\partial B_{r}}{\partial \theta}
                &\frac{1}{\sin\theta} \frac{\partial B_{r}}{\partial \phi}  \\
  \frac{\partial B_{r}}{\partial \theta}                               &0
                &0 \\
  \frac{1}{\sin\theta} \frac{\partial B_{r}}{\partial \phi}     & 0
                &0
\end{array}
\right)\right|_{{\rm PIL}^{\!\!*}}   \, .
\end{eqnarray}
The eigenvalues of this matrix are
\begin{eqnarray}
     \lambda_{1} &=& 0, \\
     \lambda_{2,3} &=& \pm \left| \nabla_{\tau} B_{r} \right|_{{\rm PIL}^{\!\!*}} \,  .
\end{eqnarray}
The eigenvector corresponding to $\lambda_{1}$ is a vector tangential to the source-surface PIL, while the remaining two eigenvectors are tangential to the helmet-streamer separatrix dome.
Each of them \fb{lays} in the plane perpendicular to the PIL at angle of $45^{\circ}$ to the local radial direction towards the Sun.
The resulting displacements along them are as follows:
\begin{eqnarray}
	\label{DrPFSS}
  \Delta r         & = &  - \frac{\Delta l }{\sqrt{2}}  \, , \\
	\label{DtPFSS}
 \Rout^{*}\, \Delta \theta & = & 
     \mp  \frac{\left. \displaystyle \frac{\partial B_{r}}{\partial \theta} \right|_{{\rm PIL}^{\!\!*}} \Delta l  }
                     { \sqrt{2}\, \Rout^{*} \left| \nabla_{\tau}\, B_{r} \right|_{{\rm PIL}^{\!\!*}} }          \, , \\
	\label{DpPFSS}
 \Rout^{*} \sin \theta_{{\rm PIL}^{\!\!*}}\, \Delta \phi    & = & 
    \mp \frac{ \left. \displaystyle \frac{1}{\sin\theta} \frac{\partial B_{r}}{\partial \phi} \right|_{{\rm PIL}^{\!\!*}} \Delta l  }
                   { \sqrt{2}\, \Rout^{*} \left| \nabla_{\tau}\, B_{r} \right|_{{\rm PIL}^{\!\!*}} }   \, ,
\end{eqnarray}
where
\begin{eqnarray}
 \Rout^{*} \left| \nabla_{\tau}\, B_{r} \right|_{{\rm PIL}^{\!\!*}} =
  \left. \left[ 
      \left(                               \frac{\partial B_{r}}{\partial\theta}\right)^{2}
    +\left( \frac{1}{\sin\theta} \frac{\partial B_{r}}{\partial\phi   }\right)^{2}
   \right]^{1/2} \right|_{{\rm PIL}^{\!\!*}} \, .
\end{eqnarray}
As for the MHD case, the upper and lower signs in equations (\ref{DtPFSS}) and (\ref{DpPFSS}) correspond, respectively,  to backward and forward displacements from the apex relative to the local magnetic field.

%
%%%%%%%%%%%%%%%%%%%%%%%%%%%%%%%%%%%%%%%%%%%%%%%%%
%%
%%                           Bibliography
%%
%%%%%%%%%%%%%%%%%%%%%%%%%%%%%%%%%%%%%%%%%%%%%%%%%

\bibliographystyle{apj}

%\bibliography{titov_bib}

\begin{thebibliography}{53}
\expandafter\ifx\csname natexlab\endcsname\relax\def\natexlab#1{#1}\fi

\bibitem[{{Altschuler} \& {Newkirk}(1969)}]{Altschuler1969}
{Altschuler}, M.~D., \& {Newkirk}, G. 1969, \solphys, 9, 131

\bibitem[{Antiochos {et~al.}(2011)Antiochos, Miki{\'c}, Titov, Lionello, \&
  Linker}]{Antiochos2011}
Antiochos, S.~K., Miki{\'c}, Z., Titov, V.~S., Lionello, R., \& Linker, J.~A.
  2011, \apj, 731, 112

\bibitem[{{Aulanier} {et~al.}(2005){Aulanier}, {Pariat}, \&
  {D{\'e}moulin}}]{Aulanier2005}
{Aulanier}, G., {Pariat}, E., \& {D{\'e}moulin}, P. 2005, \aap, 444, 961

\bibitem[{{Bungey} {et~al.}(1996){Bungey}, {Titov}, \& {Priest}}]{Bungey1996}
{Bungey}, T.~N., {Titov}, V.~S., \& {Priest}, E.~R. 1996, \aap, 308, 233

\bibitem[{{Cane} {et~al.}(2006){Cane}, {Mewaldt}, {Cohen}, \& {von
  Rosenvinge}}]{Cane2006}
{Cane}, H.~V., {Mewaldt}, R.~A., {Cohen}, C.~M.~S., \& {von Rosenvinge}, T.~T.
  2006, \jgr, 111, A06S90

\bibitem[{{Crooker} {et~al.}(2002){Crooker}, {Gosling}, \&
  {Kahler}}]{Crooker2002}
{Crooker}, N.~U., {Gosling}, J.~T., \& {Kahler}, S.~W. 2002, \jgr, 107, 1028

\bibitem[{{DeForest} {et~al.}(2012){DeForest}, {Howard}, \&
  {McComas}}]{DeForest2012}
{DeForest}, C.~E., {Howard}, T.~A., \& {McComas}, D.~J. 2012, \apj, 745, 36

\bibitem[{{D{\'e}moulin} {et~al.}(1996){D{\'e}moulin}, {Henoux}, {Priest}, \&
  {Mandrini}}]{Demoulin1996}
{D{\'e}moulin}, P., {Henoux}, J.~C., {Priest}, E.~R., \& {Mandrini}, C.~H.
  1996, \aap, 308, 643

\bibitem[{Desai \& Giacalone(2016)}]{Desai2016}
Desai, M., \& Giacalone, J. 2016, Living Reviews in Solar Physics, 13, 3

\bibitem[{Haynes \& Parnell(2010)}]{Haynes2010}
Haynes, A.~L., \& Parnell, C.~E. 2010, Physics of Plasmas, 17, 092903

\bibitem[{{Jacques}(1977)}]{Jacques1977}
{Jacques}, S.~A. 1977, \apj, 215, 942

\bibitem[{{Jin} {et~al.}(2016){Jin}, {Schrijver}, {Cheung}, {DeRosa}, {Nitta},
  \& {Title}}]{Jin2016}
{Jin}, M., {Schrijver}, C.~J., {Cheung}, M.~C.~M., {DeRosa}, M.~L., {Nitta},
  N.~V., \& {Title}, A.~M. 2016, \apj, 820, 16

\bibitem[{{Lau} \& {Finn}(1990)}]{Lau1990}
{Lau}, Y.-T., \& {Finn}, J.~M. 1990, \apj, 350, 672

\bibitem[{Linker {et~al.}(2011)Linker, Lionello, Miki{\'c}, Titov, \&
  Antiochos}]{Linker2011}
Linker, J.~A., Lionello, R., Miki{\'c}, Z., Titov, V.~S., \& Antiochos, S.~K.
  2011, \apj, 731, 110

\bibitem[{{Lionello} {et~al.}(2009){Lionello}, {Linker}, \&
  {Miki{\'c}}}]{Lionello2009}
{Lionello}, R., {Linker}, J.~A., \& {Miki{\'c}}, Z. 2009, \apj, 690, 902

\bibitem[{Liu {et~al.}(2016)Liu, Kliem, Titov, Chen, Wang, Wang, Liu, Xu, \&
  Wiegelmann}]{Liu2016}
Liu, R. {et~al.} 2016, \apj, 818, 148

\bibitem[{{Liu}(2007)}]{Liu2007b}
{Liu}, Y. 2007, \apjl, 654, L171

\bibitem[{{Liu} \& {Hayashi}(2006)}]{Liu2006}
{Liu}, Y., \& {Hayashi}, K. 2006, \apj, 640, 1135

\bibitem[{{Masson} {et~al.}(2013){Masson}, {Antiochos}, \&
  {DeVore}}]{Masson2013}
{Masson}, S., {Antiochos}, S.~K., \& {DeVore}, C.~R. 2013, \apj, 771, 82

\bibitem[{Masson {et~al.}(2009)Masson, Pariat, Aulanier, , \&
  Schrijver}]{Masson2009}
Masson, S., Pariat, E., Aulanier, G., , \& Schrijver, C.~J. 2009, \apj, 700,
  559

\bibitem[{{Miki{\'c}} {et~al.}(2013){Miki{\'c}}, {Lionello}, {Mok}, {Linker},
  \& {Winebarger}}]{Mikic2013c}
{Miki{\'c}}, Z., {Lionello}, R., {Mok}, Y., {Linker}, J.~A., \& {Winebarger},
  A.~R. 2013, \apj, 773, 94

\bibitem[{{Mok} {et~al.}(2016){Mok}, {Miki{\'c}}, {Lionello}, {Downs}, \&
  {Linker}}]{Mok2016}
{Mok}, Y., {Miki{\'c}}, Z., {Lionello}, R., {Downs}, C., \& {Linker}, J.~A.
  2016, \apj, 817, 15

\bibitem[{{Pariat} \& {D{\'e}moulin}(2012)}]{Pariat2012}
{Pariat}, E., \& {D{\'e}moulin}, P. 2012, \aap, 541, A78

\bibitem[{Parnell {et~al.}(2010)Parnell, Haynes, \& Galsgaard}]{Parnell2010}
Parnell, C.~E., Haynes, A.~L., \& Galsgaard, K. 2010, \jgr, 115, A02102

\bibitem[{Parnell {et~al.}(1996)Parnell, Smith, Neukirch, \&
  Priest}]{Parnell1996}
Parnell, C.~E., Smith, J.~M., Neukirch, T., \& Priest, E.~R. 1996, Physics of
  Plasmas, 3, 759

\bibitem[{{Priest} \& {D{\'e}moulin}(1995)}]{Priest1995}
{Priest}, E.~R., \& {D{\'e}moulin}, P. 1995, \jgr, 100, 23443

\bibitem[{{Priest} \& {Forbes}(2000)}]{Priest2000}
{Priest}, E.~R., \& {Forbes}, T.~G. 2000, {Magnetic Reconnection: MHD theory
  and applications} (Cambridge, UK: Camb. University Press, 612 p.)

\bibitem[{{Priest} \& {Titov}(1996)}]{Priest1996}
{Priest}, E.~R., \& {Titov}, V.~S. 1996, Philos. Trans. R. Soc. London A, 354,
  2951

\bibitem[{Reames(2002)}]{Reames2002}
Reames, D.~V. 2002, \apjl, 571, L63

\bibitem[{{Reames}(2013)}]{Reames2013}
{Reames}, D.~V. 2013, \ssr, 175, 53

\bibitem[{{Riley} {et~al.}(2006){Riley}, {Linker}, {Miki{\'c}}, {Lionello},
  {Ledvina}, \& {Luhmann}}]{Riley2006}
{Riley}, P., {Linker}, J.~A., {Miki{\'c}}, Z., {Lionello}, R., {Ledvina},
  S.~A., \& {Luhmann}, J.~G. 2006, \apj, 653, 1510

\bibitem[{{Ru{\v s}in} {et~al.}(2010){Ru{\v s}in}, {Druckm{\"u}ller}, {Aniol},
  {Minarovjech}, {Saniga}, {Miki{\'c}}, {Linker}, {Lionello}, {Riley}, \&
  {Titov}}]{Rusin2010}
{Ru{\v s}in}, V. {et~al.} 2010, \aap, 513, A45

\bibitem[{{Savcheva} {et~al.}(2012{\natexlab{a}}){Savcheva}, {Pariat}, {van
  Ballegooijen}, {Aulanier}, \& {DeLuca}}]{Savcheva2012b}
{Savcheva}, A., {Pariat}, E., {van Ballegooijen}, A., {Aulanier}, G., \&
  {DeLuca}, E. 2012{\natexlab{a}}, \apj, 750, 15

\bibitem[{{Savcheva} {et~al.}(2012{\natexlab{b}}){Savcheva}, {van
  Ballegooijen}, \& {DeLuca}}]{Savcheva2012}
{Savcheva}, A.~S., {van Ballegooijen}, A.~A., \& {DeLuca}, E.~E.
  2012{\natexlab{b}}, \apj, 744, 78

\bibitem[{{Schatten} {et~al.}(1969){Schatten}, {Wilcox}, \&
  {Ness}}]{Schatten1969}
{Schatten}, K.~H., {Wilcox}, J.~M., \& {Ness}, N.~F. 1969, \solphys, 6, 442

\bibitem[{{Scherrer} {et~al.}(1995){Scherrer}, {Bogart}, {Bush}, {Hoeksema},
  {Kosovichev}, {Schou}, {Rosenberg}, {Springer}, {Tarbell}, {Title},
  {Wolfson}, {Zayer}, \& {MDI Engineering Team}}]{Scherrer1995}
{Scherrer}, P.~H. {et~al.} 1995, \solphys, 162, 129

\bibitem[{Schrijver \& Title(2011)}]{Schrijver2011}
Schrijver, C.~J., \& Title, A.~M. 2011, J. Geophys. Res., 116, A04108

\bibitem[{{Schrijver} {et~al.}(2013){Schrijver}, {Title}, {Yeates}, \&
  {DeRosa}}]{Schrijver2013}
{Schrijver}, C.~J., {Title}, A.~M., {Yeates}, A.~R., \& {DeRosa}, M.~L. 2013,
  \apj, 773, 93

\bibitem[{{Schwadron} {et~al.}(2015){Schwadron}, {Lee}, {Gorby}, {Lugaz},
  {Spence}, {Desai}, {T{\"o}r{\"o}k}, {Downs}, {Linker}, {Lionello},
  {Miki{\'c}}, {Riley}, {Giacalone}, {Jokipii}, {Kota}, \&
  {Kozarev}}]{Schwadron2015}
{Schwadron}, N.~A. {et~al.} 2015, \apj, 810, 97

\bibitem[{{Seehafer}(1986)}]{Seehafer1986a}
{Seehafer}, N. 1986, \solphys, 105, 223

\bibitem[{{Sheeley} {et~al.}(2009){Sheeley}, {Lee}, {Casto}, {Wang}, \&
  {Rich}}]{Sheeley2009}
{Sheeley}, Jr., N.~R., {Lee}, D.~D.-H., {Casto}, K.~P., {Wang}, Y.-M., \&
  {Rich}, N.~B. 2009, \apj, 694, 1471

\bibitem[{{Tassev} \& {Savcheva}(2016)}]{Tassev2016}
{Tassev}, S., \& {Savcheva}, A. 2016, ArXiv e-prints, 1609.00724

\bibitem[{{Titov}(2007)}]{Titov2007a}
{Titov}, V.~S. 2007, \apj, 660, 863

\bibitem[{{Titov} {et~al.}(1999){Titov}, {D{\'e}moulin}, \&
  {Hornig}}]{Titov1999a}
{Titov}, V.~S., {D{\'e}moulin}, P., \& {Hornig}, G. 1999, in ESA SP-448:
  Magnetic Fields and Solar Processes, ed. A.~{Wilson} \& {et al.}, 715--722

\bibitem[{Titov {et~al.}(2009)Titov, Forbes, Priest, Miki\'{c}, \&
  Linker}]{Titov2009}
Titov, V.~S., Forbes, T.~G., Priest, E.~R., Miki\'{c}, Z., \& Linker, J.~A.
  2009, \apj, 693, 1029

\bibitem[{{Titov} {et~al.}(2002){Titov}, {Hornig}, \&
  {D{\'e}moulin}}]{Titov2002}
{Titov}, V.~S., {Hornig}, G., \& {D{\'e}moulin}, P. 2002, \jgr, 107, 1164

\bibitem[{{Titov} {et~al.}(2008){Titov}, {Miki{\'c}}, {Linker}, \&
  {Lionello}}]{Titov2008a}
{Titov}, V.~S., {Miki{\'c}}, Z., {Linker}, J.~A., \& {Lionello}, R. 2008, \apj,
  675, 1614

\bibitem[{Titov {et~al.}(2011)Titov, Miki{\'c}, Linker, Lionello, \&
  Antiochos}]{Titov2011}
Titov, V.~S., Miki{\'c}, Z., Linker, J.~A., Lionello, R., \& Antiochos, S.~K.
  2011, ApJ, 731, 111

\bibitem[{Titov {et~al.}(2012)Titov, Mikic, T{\"o}r{\"o}k, Linker, \&
  Panasenco}]{Titov2012}
Titov, V.~S., Mikic, Z., T{\"o}r{\"o}k, T., Linker, J.~A., \& Panasenco, O.
  2012, \apj, 759, 70

\bibitem[{{Titov} {et~al.}(1993){Titov}, {Priest}, \&
  {D\'{e}moulin}}]{Titov1993}
{Titov}, V.~S., {Priest}, E.~R., \& {D\'{e}moulin}, P. 1993, \aap, 276, 564

\bibitem[{{T{\"o}r{\"o}k} {et~al.}(2011){T{\"o}r{\"o}k}, {Panasenco}, {Titov},
  {Miki{\'c}}, {Reeves}, {Velli}, {Linker}, \& {De Toma}}]{Torok2011}
{T{\"o}r{\"o}k}, T., {Panasenco}, O., {Titov}, V.~S., {Miki{\'c}}, Z.,
  {Reeves}, K.~K., {Velli}, M., {Linker}, J.~A., \& {De Toma}, G. 2011, \apjl,
  739, L63

\bibitem[{{Wyper} \& {Pontin}(2014{\natexlab{a}})}]{Wyper2014b}
{Wyper}, P.~F., \& {Pontin}, D.~I. 2014{\natexlab{a}}, Physics of Plasmas, 21,
  102102

\bibitem[{{Wyper} \& {Pontin}(2014{\natexlab{b}})}]{Wyper2014a}
---. 2014{\natexlab{b}}, Physics of Plasmas, 21, 082114

\end{thebibliography}

%\begin{comment}

%\end{comment}

\clearpage

%%%%%%%%%%%%%%%%%%%%%%%%%%%%%%%%%%%%%%%%%%%%%%%%%%
%%%
%%%                           Figures
%%%
%%%%%%%%%%%%%%%%%%%%%%%%%%%%%%%%%%%%%%%%%%%%%%%%%%
%%%%%%%%%%%%%%%%%%%%%%% fig. 1 (Br+NL+CHs , CHs_MHD_vs_PFSS) %%%%%%%%

%%%%%%%%%%%%%%%%%%%%%%% fig. 2 (slogQ+CHs_MHD_vs_PFSS.eps) %%%%%%%%%%%%%%%

%%%%%%%%%%%%%%%%%%%%%%% fig. ?? (CH_conn_map_PFSS) %%%%%%%%%%%%%%%

%%%%%%%%%%%%%%%%%%%%%%% fig. ?? (DFregs_evol+Br+NL) %%%%%%%%%%%%%%%

%%%%%%%%%%%%%%%%%%%%%%% fig. ?? (3Dview_DFSSs+SCs) %%%%%%%%%%%%%%%
\begin{figure*}[ht]
%  \begin{comment}
\epsscale{1.1}
%\plotone{fr_afuh}\\
\plotone{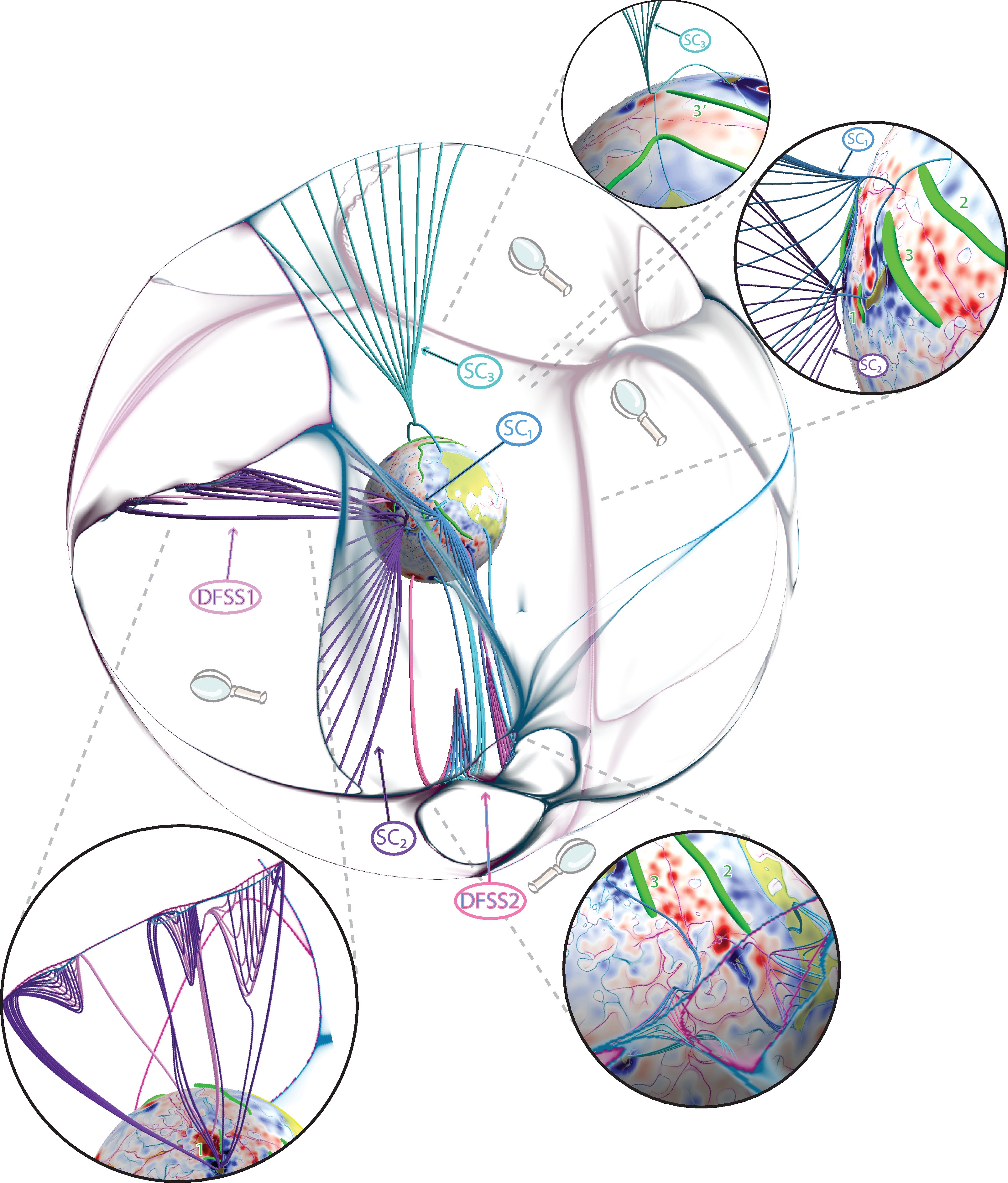}\\
%   \end{comment}
%\ \\
\caption
{
Field-line structure shown at time instant 15 in regions of interest, including the disconnected-flux separatrix surfaces DFSS1 and DFSS2, and open-field parts of separatrix curtains \SC{1\ldots3}. 
The $\slog Q$ map at the upper boundary $r=\Rout^*(\approx 5\,R_{\sun})$ is made transparent for low values of $Q$ to visualize simultaneously the indicated features and the photospheric distributions of $B_{r}$ (blue-red), $\slog Q$ (aqua-crimson) and coronal holes (yellow).
 The insets present zoomed-in views of the features.
}
	\label{f:DFSSvsSC}
\end{figure*}

%%%%%%%%%%%%%%%%%%%%%%% fig. ?? (s_blob) %%%%%%%%%%%%%%%
\begin{figure*}[ht]
%  \begin{comment}
\epsscale{1.0}
%\plotone{fr_afuh}\\
\plotone{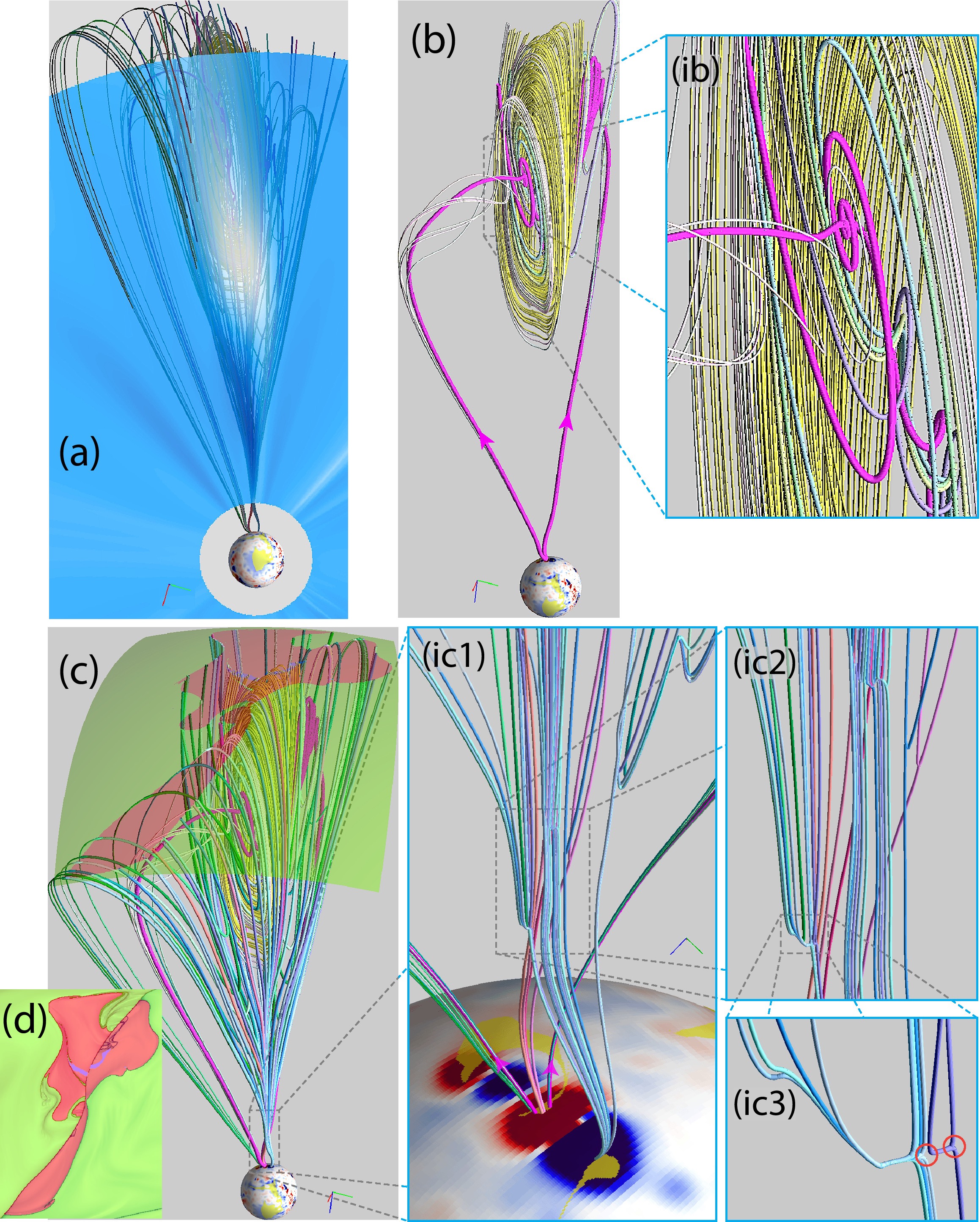}\\
%   \end{comment}
%\ \\
\caption
{
The magnetic topology of the transient DFR (at time instant 11) encircled by the white dashed line in Figure \ref{f:DFregs_ev}: (a) synthetic image of Thompson scattering brightness using a semi-transparent blue-white palette and magnetic field lines passing through a plasma blob (white spot) moving outward from the Sun\tablenotemark{b};
%\footnote{Note that the view is exactly onto the northern pole of the Sun, so the brightness image would be differentif the integration were made along the viewing direction of this figure.};
the magnetic field topology inside (b) and at the boundary (c) of the DFR, enclosing the blob, and open-disconnected (green-brown) flux areas at the upper boundary $r=20\, R_{\sun}$; (d) the same mask blended with the respective $\slog Q$ map.
Insets to panels (b) and (c) present zoomed views of the field structure near null points: a spiral one (ib) and
several radial ones (ic1--ic3); semi-transparent yellow shadings over the photospheric $B_{r}$-map mark open-field regions in inset (ic1); the red circles in inset (ic3) localize two nulls at a separator field line.\\
\\
{\footnotesize $^{\rm b}$\ Note that the view is exactly onto the northern pole of the Sun, so the brightness image would be different if the integration were made along the viewing direction of this figure.}
}
	\label{f:s_blob}
\end{figure*}
%\footnotetext{Note that the view is exactly onto the northern pole of the Sun, so the brightness image would be different if the integration were made along the viewing direction of this figure.}

%%%%%%%%%%%%%%%%%%%%%%% fig. ?? (SC1) %%%%%%%%%%%%%%%
\begin{figure*}[ht]
%  \begin{comment}
\epsscale{1.1}
%\plotone{fr_afuh}\\
\plotone{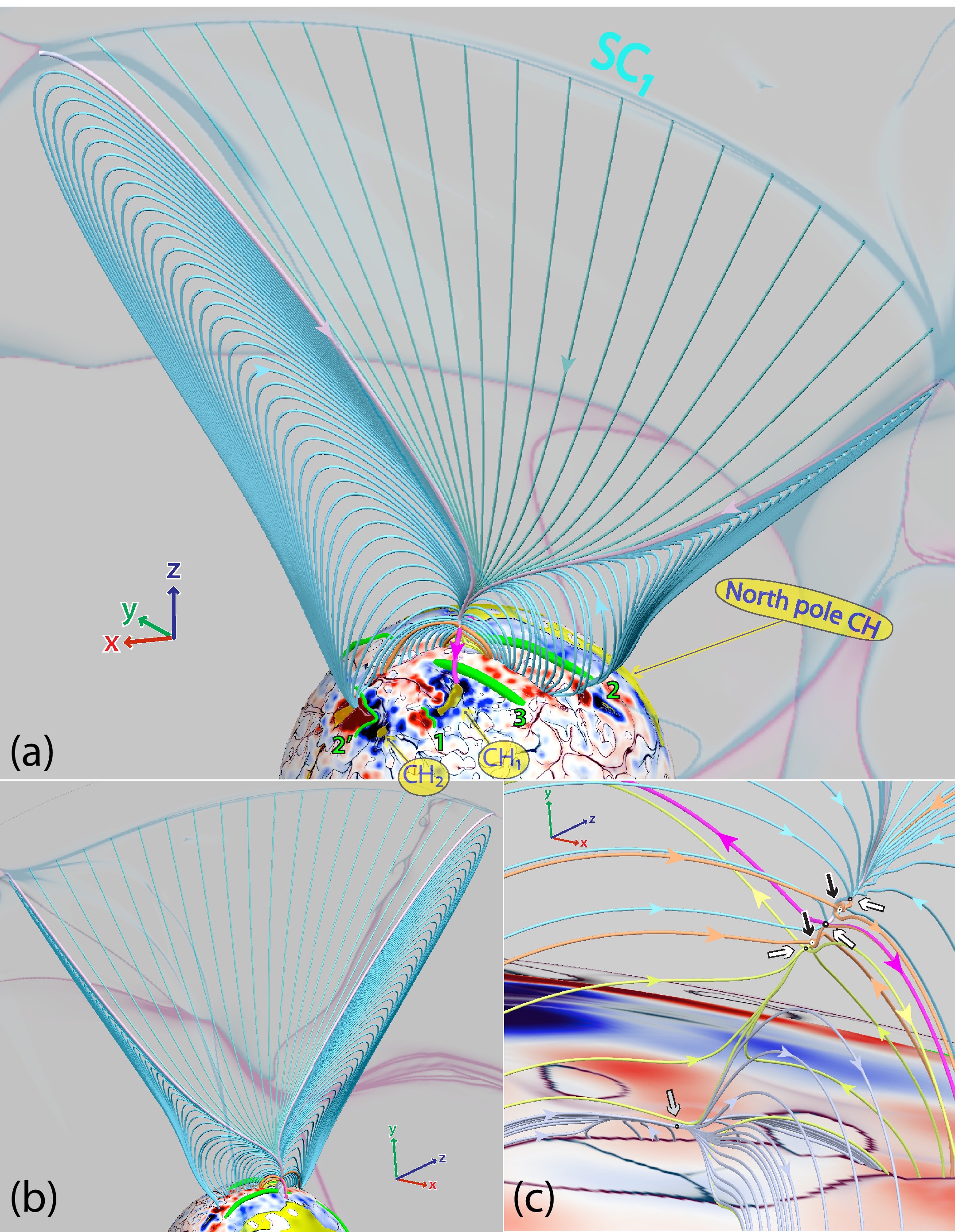}\\
%   \end{comment}
%\ \\
\caption
{
Separatrix curtain \SC{1} at the final instant 15 of the MHD run: a view on \SC{1} similar to the one used for the PFSS model 
\cite[(a), cf. Figure 5 in][]{Titov2012}, a view on the other side of \SC{1} (b), and a zoomed region around the basic null point (c).
Black and white arrows indicate spiral and radial null points, respectively, while the grey arrow points to a degenerate null.
The basic null point is the one to which the spine (thick magenta) lines connect.
The vector triads on these panels show the orientation of the Cartesian system that is rigidly bound to the Sun center with the $z$-axis directed to the north pole.}
	\label{f:SC1}
\end{figure*}

\end{document}